\documentclass{article}
\usepackage{graphicx}

\usepackage{subfigure}
\usepackage{multirow}
\usepackage{appendix}
\usepackage{calc}
\usepackage{soul}
\usepackage{epsfig}
\usepackage{svg}
\usepackage{float}
\usepackage{amsmath,amssymb,amsthm}
\usepackage[a4paper,margin=1.2in]{geometry}
\usepackage{mathtools}
\usepackage{multirow}
\usepackage{mathrsfs}
\usepackage{mathtools}
\usepackage{braket}
\usepackage{amsmath}
\usepackage{lipsum}
\usepackage{siunitx}
\usepackage{physics}
\AtBeginDocument{\RenewCommandCopy\qty\SI}
\usepackage{authblk}
\usepackage{steinmetz}

\usepackage[style=numeric, sorting=none, giveninits=true,url=false]{biblatex}
\DeclareSourcemap{
  \maps[datatype=bibtex]{
    \map{
      \step[fieldset=language, null]  
      \step[fieldset=note, null]       
      \step[fieldset=issn, null]
      \step[fieldset=isbn, null]       
      \step[fieldset=abstract, null]
      \step[fieldset=file, null]
    }
  }
}
\DeclareMathOperator{\sgn}{\mathrm{sgn}}

\newtheorem{proposition}{Proposition}

\renewcommand{\theequation}{\arabic{section}.\arabic{equation}}

\makeatletter
\newcounter{pequation}
\AtBeginEnvironment{proposition}{%
\let\c@equation\c@pequation%
\def\theequation{P.\thepequation}%
\setcounter{pequation}{1000+\value{proposition}}
}

\makeatother

\title{Mechanical quality factor estimation from a nonlinear optical cavity response}

\author[1$\dagger$]{Zohran Ali\thanks{Corresponding author: zohal@dtu.dk}}
\author[1$\dagger$]{Davide Tomasella}
\author[1,2]{Daniel Allepuz-Requena}
\author[1]{Alexander Huck}
\author[1]{Ulrik Lund Andersen}
\affil[1]{Center for Macroscopic Quantum States (bigQ), Department of Physics,
Technical University of Denmark, 2800 Kongens Lyngby, Denmark}
\affil[2]{James C. Wyant College of Optical Sciences, The University of Arizona, Tucson, AZ 85721, USA}

\AtBeginDocument{\RenewCommandCopy\qty\SI}

\begin{document}

\maketitle
\let\thefootnote\relax\footnotetext{$\dagger$ These authors contributed equally to this work.}
\vspace*{-3em}
\begin{abstract}

Intrinsic damping rate and quality factor ($Q$) are key parameters in optomechanical systems. However, in high-Q devices, their measurement is often challenging because probe-induced dynamical backaction (DBA) and the need to resolve and accurately track the mechanical resonance limit the performance of state-of-the-art techniques. 
Here, we introduce an all-optical ringdown method in the unresolved-sideband regime that exploits the nonlinear, time-averaged optical response induced by strong mechanical motion. 
We derive an analytical model of the time-averaged optical cavity response probed by a weak frequency-scanning laser. During the evolution of the mechanical oscillator, the excited motion modulates the Lorentzian lineshape into a double-horned profile whose peak separation enables an accurate estimation of the modulation amplitude. Tracking this response during the mechanical ringdown allows the intrinsic damping rate to be determined without resolving the mechanical oscillation.
We apply the method to a density-modulated phononic crystal membrane placed inside a high-finesse cavity and obtain mechanical quality factors consistent with independent measurements on the same platform.
Our method requires a minimal optical setup, while offering an in situ diagnostic and reduced impact from probe-induced DBA and robustness against measurement noise. 
Finally, the model can describe other resonant systems dispersively coupled to coherent frequency modulations, providing new insights and measurement strategies that expand beyond the optomechanical platform.

\end{abstract}

\begin{refsection}
\counterwithout{equation}{section}
\section{Introduction}

A great variety of physical degrees of freedom can be measured through dispersive coupling to a resonant system. 
Resonators improve measurement sensitivity and signal-to-noise ratio (SNR) \cite{Verlot2010-nu, zhang_optomechanical_2022, Reschovsky2022-eb, fait_high_2021} by converting small variations of the input signal into amplified output responses, often shifting them into more accessible frequency ranges.
Such experiments can provide a linear transduction of the input signal under the assumption that the variation is small compared to the resonance linewidth \cite{aspelmeyerCavityOptomechanics2014, Bemani2025-mr}.
\par
However, when fluctuations become comparable to or exceed the resonance linewidth, the linear approximation breaks down, and the full nonlinear response of the resonant system must be taken into account \cite{clarke_cavity_2023, allepuz-requena_mitigating_2026}. Correction of the nonlinearity requires full characterization of the system's response, which is then used in an estimation protocol applied along the measurement record. Conventional reconstruction of the instantaneous nonlinear response therefore requires a detection bandwidth sufficient to resolve the modulation and a digitization rate satisfying the corresponding sampling requirements. For fast mechanical oscillations, this can impose demanding bandwidth and data-acquisition requirements.

\par
In this work, we show that the nonlinear response of a resonant system can still be used for parameter estimation when the underlying coherent modulation is too fast to be directly resolved. We derive a closed-form expression for the time-averaged response of a Lorentzian resonance subject to sinusoidal frequency modulation and analyze its behavior in both the weak- and strong-modulation regimes. Although the Lorentzian profile is progressively distorted, the averaged resonance retains quantitative information about the modulation amplitude. In the strong-modulation regime, the double-horned profile that emerges admits an asymptotically affine relation between the peak positions and the modulation amplitude.

\par
To validate our model, we applied it to an optomechanical cavity system.
By measuring the cavity's output fields, the mechanical displacement can be inferred with imprecision on the order of the mechanical zero-point fluctuations \cite{Anetsberger2009-nc}, enabling state-of-the-art performance in position and interferometric sensing \cite{Bond2016-ow, Martynov2016-dq}.
Recent theoretical \cite{clarke_cavity_2023} and experimental work \cite{allepuz-requena_mitigating_2026} have shown that such systems can enter the nonlinear regime when the mechanical oscillations are driven into large coherent states. 
In this regime, conventional reconstruction of the instantaneous mechanical displacement requires resolving and modeling the nonlinear cavity transduction over the mechanical cycle, which can impose demanding detection-bandwidth and calibration requirements. Here we show that, when only the slowly varying modulation amplitude is required, these fast dynamics need not be resolved.

\par
This provides a direct method for estimating the quality factor of mechanical resonators from the averaged spectral response, avoiding limitations associated with DBA \cite{huang_room-temperature_2024, sudhir_quantum_2018}, laser-phase-noise-induced squashing \cite{Safavi-Naeini2013-ea}, and relaxing the setup requirements for frequency calibration and stability. 
More generally, it illustrates that nonlinear distortion in dispersively coupled resonant systems does not necessarily prevent parameter estimation outside the measurement bandwidth.

\par
\vspace{\baselineskip}
In the first section, we present the model and prove the approximate result in both weak and strong modulation regimes. In the next section, we present the experimental setup and the results of applying our method to estimate the quality factor of a resonator, comparing it to other known techniques. A detailed proof and analysis of the proposed theoretical model is carried out in the Appendix section, together with a numerical model describing the time-domain dynamics qualitatively.

\section{Analytical Model }
\subsection{Analytical model of a modulated resonant system}
Resonant systems can be used as transducers, enabling variations of a second system to couple to more easily measurable physical properties. The power spectral density (PSD) of a resonant system usually follows a Lorentzian lineshape $\mathcal{L}(\omega, \omega_0)=1/(1+[2(\omega-\omega_0)/\kappa]^2)$, characterized by a linewidth $\kappa$ and a resonant frequency $\omega_0$, which can be measured by scanning a probing signal around the resonant frequency. The second system can couple to the resonant system by modifying its linewidth, referred to as dissipative coupling, or its resonant frequency, i.e., dispersive coupling. Variations of $\kappa$ or $\omega_0$ result in measurable changes---either in amplitude or phase---of the resonant response \cite{villar_conversion_2008}.
\par

In the following, we analyze the effects of a coherent modulation of the resonance frequency $\omega_1 = \omega_0 + \alpha\sin(\Omega_mt +\phi_0)$. In our example, this is the result of dispersive coupling between an optical cavity and a high-Q mechanical resonator, e.g., a phononic membrane oscillating at frequency $\Omega_m$, with $\alpha$ denoting the cavity-frequency modulation amplitude (proportional to the mechanical displacement amplitude). Therefore, measurements of the modulation strength applied to our resonant system represent measurements of the mechanical oscillation amplitude.
\par
We can distinguish two measurement regimes based on the response of our measurement apparatus. If the detection bandwidth encompasses the mechanical frequency and the signal is sampled sufficiently rapidly, the instantaneous mechanical oscillation can be resolved. Here we instead consider the opposite limit, $\mathrm{BW}\ll\Omega_m/2\pi$, in which the detector records only the response averaged over many mechanical periods. For the optomechanical cavity considered in this work, we operate in the unresolved-sideband regime, $\Omega_m\ll\kappa$, such that the intracavity field follows the mechanical frequency modulation adiabatically. \\
Under these conditions, the time-averaged signal $\Lambda(\omega,\alpha)$ can be described by
\begin{equation}
    \Lambda(\omega,\alpha)=\frac{1}{T}\int_{0}^{T}\mathcal{L}(\omega,\omega_0+\alpha\sin(\Omega_mt +\phi_0))\dd{t},
    \label{eq:integral_lorent}
\end{equation}
where $T=1/\mathrm{BW}$ is the measurement time.

After normalizing the detuning ${\nu=2(\omega-\omega_0)/\kappa}$ and the amplitude ${A=2\alpha/\kappa}$ with the resonance's Half-Width Half-Maximum (${\mathrm{HWHM}=\kappa/2}$), the integral is rewritten in a more general form

\begin{equation}
    \Lambda(\nu,A)=\frac{1}{2\pi}\int_{0}^{2\pi}\frac{\dd{\phi}}{1+\left(\nu+A\sin\phi\right)^2}.
    \label{eq:integral_period_phi}
\end{equation}
\par
This form of integral has appeared in different contexts in the literature \cite{arcizet_single_2011, Castignani2023-ys} with numerical solutions tailored to specific applications.
Analytical studies of lineshape broadening for Lorentzian peaks have also been previously reported in \cite{Arndt1965-dq, Axner2001-ul, Mei2015-jr} for their role in wavelength modulation spectroscopy (WMS) and Fourier component analysis.
Here, we derive the time-averaged response, which coincides with the zeroth-order Fourier component of the expressions in \cite{Arndt1965-dq, Axner2001-ul}, and express it in a compact form suitable for the present problem.
Using complex analysis and contour integration, we obtain the closed-form, real-valued expression for the solution as
\begin{equation}
\Lambda(\nu,A)=\sqrt{\frac{\sqrt{\eta_+\eta_-}+\eta_A}{2\eta_+\eta_-}},
\label{eq:real_solution}
\end{equation}
with the definitions $\eta_A=1+A^2-\nu^2$ and $\eta_\pm=1+(A\pm\nu)^2$. In Appendix~\ref{a:solution}, we report the mathematical derivation.
This formulation provides deeper insight into the system's response across different modulation-strength regimes. To our knowledge, the extrema of this response have not been analyzed; in the following, we derive the condition for the maxima, the splitting threshold $A=\sqrt2$, and the exact and asymptotic peak positions, which form the basis of the amplitude estimator used in this work.

\begin{figure}[ht]
    \centering
    \includegraphics[width=\textwidth]{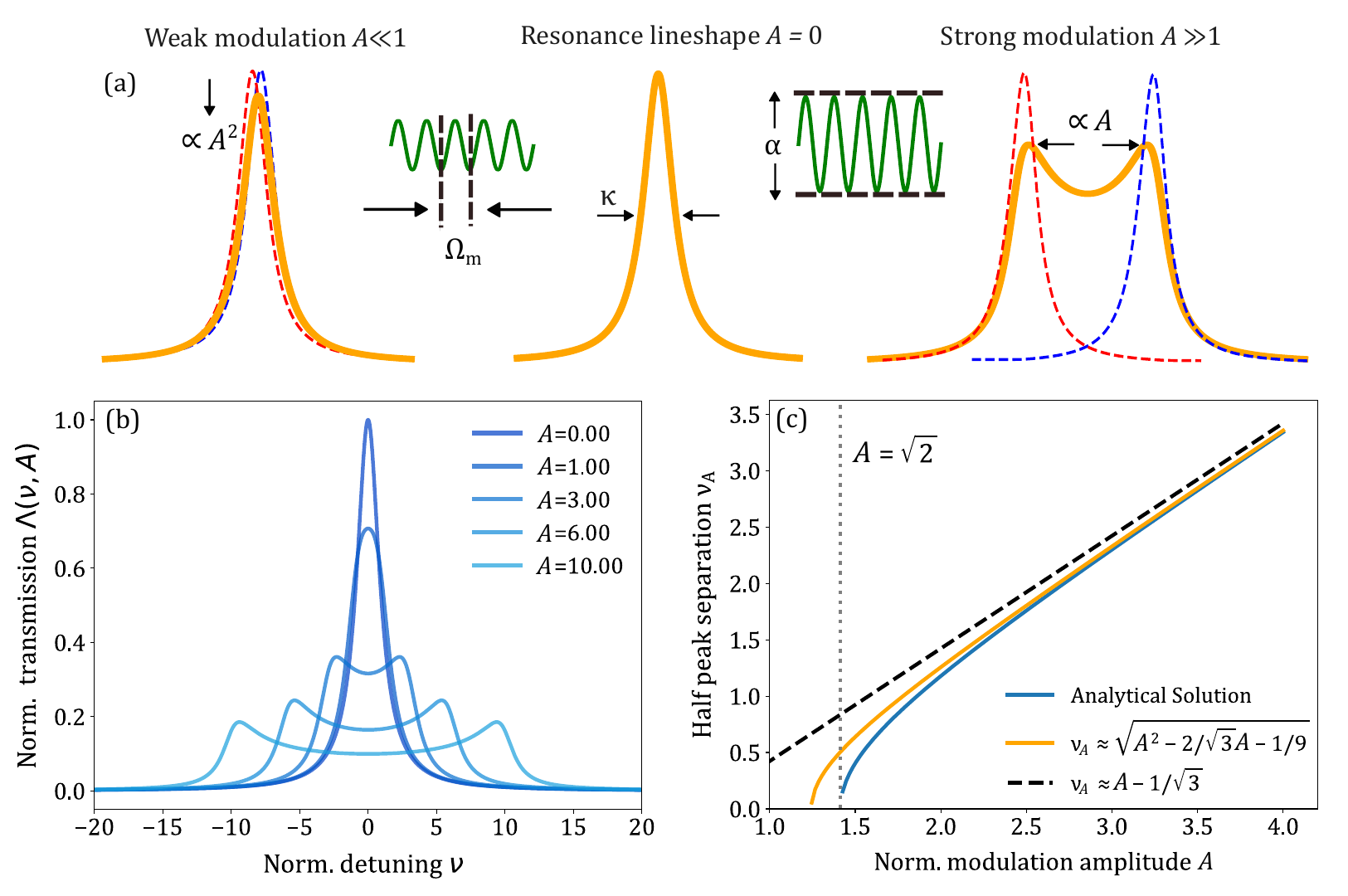}
    \caption{
       (a) Illustration of the effect of different modulation strengths $\alpha$ on the averaged lineshape profile. The coherent drive with high frequency $\Omega_m$ modulates the center frequency of the resonant system (red and blue dashed lines) in the dispersive coupling regime. On the left, when the normalized modulation amplitude is weak ($A\ll1$), the Lorentzian peak's height is reduced proportionally to $A^2$. On the right, when we have a strong modulation ($A\gg1$), we enter the nonlinear regime, and the measured resonance takes a double-horned profile. (b) Model for the slow measurement of the resonant system as a function of the normalized detuning $\nu$ for different modulation strengths $A$. (c) Comparison between the analytical solution for the peak separation (blue) and the approximated solution discussed in the text (black and orange). The asymptotic behavior $\nu_A=A-1/\sqrt3$ shows that, in the strong-modulation regime, the peak position is approximately affine in the modulation amplitude, with a constant offset. The vertical line shows the threshold $A=\sqrt{2}$ for the splitting of the resonance.
    }
    \label{fig:peak_split_model}
\end{figure}

\subsection{Amplitude estimation of a fast coherent modulation}

An analysis of the solution in Equation~\ref{eq:real_solution} allows us to identify two regimes---weak and strong modulation---depending on the ratio ($A$) between modulation amplitude ($\alpha$) and the resonance's linewidth ($\kappa$), as shown in Figure~\ref{fig:peak_split_model}(a).
While the strong modulation regime requires further analysis, the solution with a weak coherent drive ($A\ll1$) has a simple Lorentzian approximation across the entire spectrum:
\begin{equation}\label{eq:smallv_claim}
\Lambda(\nu,A)\simeq\frac{1-\frac{A^2}{2}}{1+\nu^2}+O\left(A^4+\min\left(A^2\nu^2+\nu^4, \frac{A^2}{\nu^2}+\frac{1}{\nu^4}\right)\right), \quad\forall \nu.
\end{equation}
The derivation of this result from our general model is presented in Appendix~\ref{a:weak_approx}.
When the normalized modulation amplitude is small, the frequency response is still a Lorentzian peak with the same linewidth and a reduced height.
We can extract information about the modulation amplitude from the change in the peak height $\Delta h$ with an inverse quadratic relation $A\sim\sqrt{2\Delta h}$.
This shows that the sensitivity $S=\partial{\Lambda}/\partial{A}=-A/(1+\nu^2)$ is maximized by measuring the variation of the in-resonance response ($\omega=\omega_0$), in contrast to the result for high-bandwidth detection in which the best sensitivity is obtained when the slope of the transfer function is maximized.
\\
When $A=0$, we also verify that both the full solution and the approximation converge to the initial normalized Lorentzian peak.

\par
With an increase in the modulation amplitude, the resonance's response changes from an approximated Lorentzian to a double-horned profile, as shown in Figure~\ref{fig:peak_split_model}(a-b). As a result, the single-peak-height estimator is no longer the most natural descriptor for the modulation amplitude. However, a more robust approach is possible for strong modulation drives ($A\gg1$), now exploiting the separation between the two off-center maxima $2\nu_A$.

For that, starting from our previous expression in Equation~\ref{eq:real_solution}, we can evaluate the derivative of the response and find the critical points $\partial{\Lambda}/\partial{\nu}=0$ corresponding to the profile maxima as
\begin{equation}
\eta_\nu^3+\left(\frac{3}{2} A^2-2\right) \eta_\nu^2-4 A^2 \eta_\nu-2 A^4=0,
\label{eq:derivative_solution}
\end{equation}
where $ \eta_{\nu} = 1- A^2 +\nu^2$. In Appendix~\ref{a:strong_distance}, we prove this result and show that the solution for the maxima is unique and the only one that satisfies the condition $0<\nu_A<A$.
Another useful result is derived in Appendix~\ref{a:strong_approx}, where we prove the following approximation for $A\gg1$:
\begin{equation}\label{eq:approximation}
\nu_A\simeq\sqrt{A^2-\frac{2}{\sqrt{3}}A-\frac{1}{9}+O\left(A^{-1}\right)}\simeq A-\frac{1}{\sqrt{3}}+O\left(A^{-1}\right).
\end{equation}
This result follows the expected limit for $A\gg1$ with a linear increase in peak separation for large modulation $A\sim\nu_A+1/\sqrt{3}$. Hence, the peak position is asymptotically affine in the modulation amplitude, with a constant offset $1/\sqrt{3}$. Moreover, our approach provides an analytical proof for the constant offset of the asymptote and a better quadratic approximation for smaller modulation amplitudes.
\par
Other meaningful considerations can be extrapolated from our model. In Figure~\ref{fig:peak_split_model}(c), we plot a comparison between the accuracy of the analytical solution and its approximation as a function of the modulation amplitude $A$.
For this result, the sensitivity $\partial{\Lambda}/\partial{A}$ does not have a direct analytical form since it depends on how the peak separation is determined with a frequency scan across the peak frequency.
Furthermore, we also determined the threshold ($A=\sqrt{2}$) for the splitting of the averaged resonance response, as shown in Appendix~\ref{a:strong_distance}. Finally, a closed-form analytical model enables fitting of experimental data even below this threshold and provides insight into the dynamics across the full modulation amplitude range.

\section{Experimental results}

\subsection{Ringdown measurements of a mechanical resonator}

The natural damping rate $\Gamma_m$ of a mechanical resonator is one of the parameters to be characterized in cavity optomechanics experiments. At room temperature, $\Gamma_m$ determines the thermal decoherence rate and thus the feasibility of quantum regime experiments such as ponderomotive squeezing~\cite{huang_room-temperature_2024} or ground-state preparation~\cite{pluchar_towards_2020}. The damping rate of a resonator can be estimated through a ringdown measurement: the mechanical population is initialized in a large coherent state, then the resonator is left to freely relax while measuring its amplitude. If no additional damping or anti-damping mechanisms affect the dynamics, the rate at which it relaxes will be the natural mechanical damping rate.
\par

However, the damping rate of a resonator coupled to an optical cavity is often modified due to a delayed position-dependent radiation pressure force, via DBA \cite{aspelmeyerCavityOptomechanics2014}. DBA establishes an energy transfer between the mechanical and optical fields, allowing the resonator to lose (receive) energy to (from) the optical field, resulting in additional damping (anti-damping). In dispersively coupled systems, DBA will damp the resonator when the cavity is driven red-detuned ($\nu<0$) and will anti-damp when blue-detuned ($\nu>0$). The DBA vanishes when the cavity is driven on resonance ($\nu=0$). Thus, only when the cavity is driven on resonance will a ringdown yield the natural damping rate.
\par
Quantum optomechanical experiments are engineered to achieve low natural damping rates and large optomechanical interactions, meaning that even at a small deviation from $\nu=0$, the DBA damping rate quickly overwhelms the natural one. Recent high-cooperativity experiments have addressed this issue by performing a series of ringdown measurements at decreasing probe powers and extrapolating the natural damping rate to zero power~\cite{huang_room-temperature_2024}, or by measuring the resonator before incorporating it into the optical cavity~\cite{saarinen_laser_2023}. However, both have several limitations in practice: the first approach yields a decreasing signal-to-noise ratio as probe power decreases; the second assumes that integrating the resonator into the cavity introduces no additional mechanical loss mechanisms. Moreover, the second approach might be impossible on certain experimental platforms such as optomechanical crystals, where the mechanical and optical resonators are inseparable. 
Furthermore, since high laser probe power is often required to have a significant SNR, this also worsens the DBA effect on the measurement.
\par
In addition to the DBA, other limitations arise from the technical challenges of experimentally measuring the mechanical amplitude.
For example, ringdown measurements often involve the use of a signal generator and amplitude modulator to excite the mechanical mode and then tracking the height of the mechanical peak with an electrical spectrum analyzer \cite{Kuhn2011-mm, hoj_development_2021}. This requires careful frequency calibration and precise tuning to match the mechanical frequency. For high-Q resonators, this is particularly challenging because of the narrow mechanical linewidth, requiring millihertz-level frequency resolution. Moreover, in narrow-linewidth cavities, laser phase-noise-induced squashing of the mechanical peak height could also impact the measurement performance. 
\vspace{\baselineskip}
\par
Our method provides an alternative approach to performing mechanical ringdown measurements that both simplifies the experimental requirements and addresses dynamical backaction (DBA) effects.
\par\begin{figure}[ht]
\centering
  
     \includegraphics[width=1\textwidth]{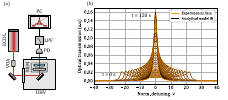}
 
     \caption{(a) Experimental setup for the measurement of the resonator's quality factor with a scanning probe laser. 
     A MIM setup comprising a $73\text{-}\mu{m}$ cavity and a membrane is placed inside the ultra-high-vacuum (UHV) chamber (gray box), and an ECDL is scanned to probe the system, with a variable optical attenuator (VOA) to adjust the laser power. The transmission of the MIM setup is measured with a photodetector (PD), and a low-pass filter (LPF) is placed on the electrical path
    between PD and the PC. The dynamics are recorded with an oscilloscope on a standard computer (PC).
     (b) The evolution of the optical resonance shape of the MIM cavity is presented. The initial state at $t =0 \:\mathrm{s}$  corresponds to the point where the laser is briefly locked on the blue side of the optical resonance, resulting in a double-horned optical lineshape. In orange, some of the measurements are plotted as a function of the normalized detuning $\nu$ for different times between $t =0 \:\mathrm{s}$ and $ t = 120 \:\mathrm{s}$. The black line is the model fit for fixed linewidth $\kappa/2\pi = 37.8\,\mathrm{MHz} $, and a free factor for the maximum amplitude.}
\label{fig:Quality_facor}
\end{figure}
After exciting the mechanics with a blue-detuned laser, the measurement is performed by continuously scanning the laser across the optical resonance.
The transmitted or reflected signal is directly measured over time using a low-bandwidth photodetector. The low-bandwidth optical readout removes the need to continuously resolve the mechanical spectrum--e.g., with an interferometric detection scheme--and any associated frequency or phase-locking requirements.
In addition, both mechanical excitation and readout are achieved optically with the same laser, so no external actuators or tuning of resonant drives are required. 
\par
The measured signal contains information about the time evolution of the mechanical amplitude.
Because the probe detuning is not fixed, the DBA varies over time and depends on the scan direction--red-to-blue or blue-to-red detuning. However, the total DBA contribution for a bidirectional scan is minimized, i.e., we are both heating and cooling by the same amount.
Furthermore, the residual effect of DBA is detected and quantified by comparing the apparent mechanical decay for red-to-blue and blue-to-red scan directions. 
Finally, the low detection bandwidth allows longer signal integration and operation at lower probe powers, further reducing the effects of DBA.

\subsection{Experimental methods}
The experimental platform we used to obtain our results is a membrane-in-the-middle (MIM) system, i.e., a phononic-crystal nanomechanical resonator made of silicon nitride \cite{hoj_ultracoherent_2024} placed between two highly reflective mirrors patterned with a phononic shield \cite{allepuz-requena_mitigating_2026, Allepuz-Requena2026-fc}. In Figure~\ref{fig:Quality_facor}(a), we show a scheme of the experimental setup to expose its simplicity.
The mechanical frequency of the system is $\Omega_m/2\pi=1.134\:\mathrm{MHz}$ while the optical resonance is close to $1535\:\mathrm{nm}$ with a linewidth $\kappa/2\pi=37.8\:\mathrm{MHz}$.
\par
The cavity resonance peak is probed with a tunable Toptica CTL 1550 external cavity diode laser (ECDL). The power transmitted through the optomechanical cavity is recorded with a low-noise FEMTO photodetector (PD) operating at a bandwidth of $1\:\mathrm{kHz}$.
\par
The relevant timescale conditions are well satisfied in our experiment. Firstly, $\Omega_m/\kappa = 0.030$, placing the optomechanical system in the unresolved-sideband regime, and allowing the optical field to follow the mechanical frequency modulation adiabatically. Secondly, the mechanical oscillation frequency is $\Omega_{m}/2\pi=1.134\:\mathrm{MHz}$, while the detector's bandwidth is 1 kHz, thus $\mathrm{BW}/(\Omega_{m}/2\pi)\approx 8.8\times 10^{-4}$. As a result, the fast mechanical motion is not resolved, and only the time-averaged cavity response is measured. Finally, the mechanical amplitude ringdown occurs on a much slower timescale than a single resonance scan ($1/\Gamma_m\gg T_w$), allowing each individual scan to be treated as probing a quasi-stationary averaged lineshape.
\par
The measurement of the mechanical quality factor involves two steps: the initialization of the coherent phonon population and the probing of the cavity response during the decay of the mechanics.
A high-phonon coherent state can be created by briefly locking the laser on the blue-detuned side of the resonance. Dynamical backaction originating in the dispersive optomechanical coupling amplifies the mechanical oscillator, driving the mechanical mode into a large-amplitude self-sustained oscillation.

This drives the mechanics strongly enough to push optomechanical transduction beyond the small-amplitude approximation. These large, coherent modulations bring the system into the nonlinear regime and produce an apparent broadening and splitting of the averaged optical resonance peak. 
After unlocking the laser, we scan the laser frequency with a triangular wave with amplitude $2\ \mathrm{GHz}$ and period $T_w=100\ \mathrm{ms}$. This allows probing the double-horned resonance shape with lower effective power while the mechanics relaxes back into a thermal state, i.e., the peak's lineshape evolves back into a Lorentzian.

\subsection{Experimental estimation of mechanical quality factor with a scanning probe laser}

As discussed in the Experimental Methods section, the analytical model we derived can be applied to our optomechanical system. In general, when a mechanical oscillator is dispersively coupled to an optical resonance, the response is highly nonlinear when the oscillator is initialized in a high-phonon state; however, our analysis proved that the averaged response is smooth and contains all the information required to extrapolate the dynamics of the oscillator.
\par
The time evolution of the oscillator amplitude $A(t)$, left free to evolve, will follow an exponential decay with time constant $\tau$ related to the mechanical damping rate $\Gamma_m$ \cite{shi_topology_2024, hoj_development_2021}

\begin{equation}\label{eq:decay} A(t)=A_{0}\exp\left(-\frac{t}{\tau}\right) =A_{0}\exp\left(-\frac{\Gamma_m t}{2}\right). \end{equation}
From this expression, we can easily find the quality factor of the resonator, as it is simply $Q_m = \Omega_m/\Gamma_m$, where the amplitude decay time $\tau$ is related to the mechanical damping
rate by $ \tau={2}/{\Gamma_m}$.
\par
In Figure~\ref{fig:Quality_facor}(b), we show the evolution of the optical resonance as observed in transmission through the cavity. The orange curves represent experimental data taken at different timestamps from the start of the ringdown. These data are fitted with our analytical model (black line), showing very good agreement.
We observe that the model predicts the response for both large and smaller amplitudes of the mechanical motion.
The optical linewidth $\kappa$, used to normalize the fit parameters, was determined independently using the sideband modulation technique for different powers. 
\begin{figure}[htbp]
\centering
  \includegraphics[width=1\textwidth]{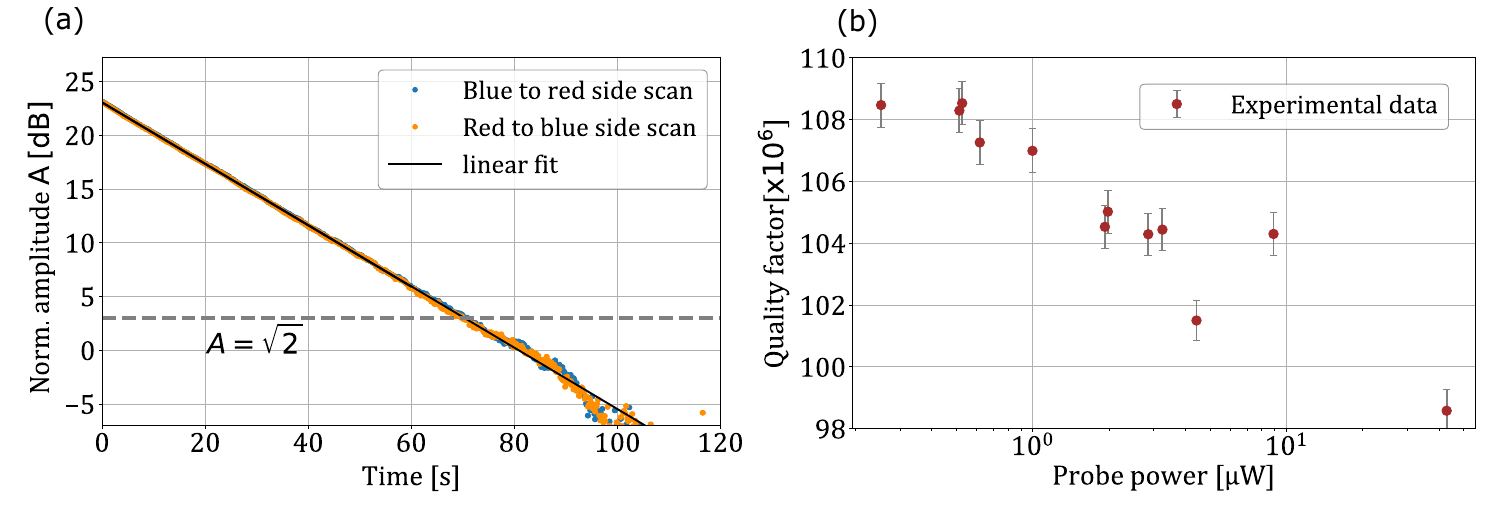}
     \caption{Experimental results for estimating the quality factor of the mechanical resonator. (a) Ringdown measurement made by estimating the peak separation from the double-horned profile. Different colors represent the scan direction; the black solid line is a linear fit over the first 60 s. From the slope of this fit, Q factors of ($108.2 \pm 0.5) \times10^6$ and ($108.2 \pm 0.5 )\times10^6$ are estimated for blue-to-red and red-to-blue detuned scans, respectively. (b) Dependence of the Q factor on the scanning power. As the scanning power decreases, the measurement of Q starts to saturate because of the negligible DBA, and this value agrees well with an independent estimate of the Q factor of the same device performed at a different time \cite{allepuz-requena_mitigating_2026}.}
\label{fig:decay_power}
\end{figure}

\par
From the extracted peak separation $2\nu_A$, the normalized amplitude $A$ can then be plotted with respect to time as in Figure~\ref{fig:decay_power}(a), allowing for the fitting of the ringdown of the mechanical motion and the extraction of the quality factor. 
From each measured resonance scan, we extract the relative modulation amplitude $A$ using the analytical model described above. For scans well above the splitting threshold, $A$ is an affine function of the peak separation, $A\simeq\nu_A+1/\sqrt3$; more generally, including lower modulation amplitudes, a value for $A$ is obtained from the fit of the full averaged lineshape.
In fact, the fitting can extract values even below the peak splitting limit ($A=\sqrt2$) until the thermal noise becomes predominant and the oscillations can no longer be described as a coherent drive. 
Plotting the inferred amplitude as a function of time yields the mechanical ringdown, from which $\Gamma_m$ and hence $Q_m$ are extracted. The decay rate is obtained from a linear fit over the first 60~s of the ringdown, where the amplitude remains above the splitting threshold.
The estimated quality factor is $(108.2\pm0.5)\times10^{6}$, in agreement within $1.5\sigma$ with the value $(107.1\pm0.5)\times10^{6}$ obtained on the same device with a far-detuned ringdown measurement performed at a different time \cite{allepuz-requena_mitigating_2026}.
\par

An important feature of our approach is that it provides an internal consistency check for probe-induced dynamical backaction. Because a blue-detuned probe heats the mechanics, whereas a red-detuned probe cools it, the effective backaction during the scan can produce different inferred amplitudes for opposite scan directions (see also Appendix~\ref{a:numerical}). For the low-probe-power data used in Figure~\ref{fig:decay_power}(a), the red-to-blue and blue-to-red scans agree within the experimental uncertainty, indicating that probe-induced backaction is negligible at that power.
In Figure~\ref{fig:decay_power}(b), at higher probe powers, we observe an additional degradation of the estimated quality factor, which may arise from optical-absorption-induced heating or thermo-optical cavity shifts \cite{Steinlechner2012-sn, Planz2023-qm}. When the scanning probe crosses the resonance, more power is coupled inside the cavity, thus heating the whole system and shifting the resonance frequency. This time-dependent shift breaks the symmetry between heating and cooling of the mechanical resonator and reduces the estimated quality factor.
\par
Another quality of our measurement technique is how the information carried by each scan evolves during the ringdown. In conventional ringdowns, a signal $S\propto A^{n}$ is measured, with $n=1$ for the oscillation amplitude and $n=2$ for the mechanical peak height in the power spectral density; for amplitude-independent noise, $\sigma_A/A\propto A^{-n}$, so the signal-to-noise ratio worsens as the oscillator decays and limits the useful part of the ringdown \cite{Brenes2025-ot}. In our method, $A$ is instead obtained from a fit of the full averaged lineshape, whose horn separation encodes the amplitude as a frequency difference. Since the lineshape area is conserved, the horn height decreases only as $A^{-1/2}$ for $A\gg1$ and, in the presence of detector noise, this yields $\sigma_A\propto A^{1/2}$. If we also take into account the fluctuations of the scan speed, which contribute $\sigma_A\propto A$, we get a rate $\sigma_A/A\propto A^{-q}$ with $0\le q\le1/2$. This bound agrees with our experimental result of $q\simeq0.2$, i.e., the information carried by each scan remains nearly constant throughout the ringdown. Moreover, as a frequency difference, $A$ is insensitive to intensity and detector-gain fluctuations and to slow drifts of the cavity resonance, making the proposed method valuable for unstable and noisy environments where conventional amplitude readouts are affected by such fluctuations.

\par
Finally, we emphasize that our technique remains unbiased with respect to the scanning direction even at relatively higher powers as long as the optical resonance remains centered within the scanning range. In fact, deviations become pronounced when the optical peak drifts because more than half the scanning period is spent on the heating/cooling side of the peak. This is another contribution to the variance of the results in Figure~\ref{fig:decay_power}(b).

\section{Conclusion}
In this paper, we have demonstrated an analytical framework for estimating the amplitude of a fast coherent modulation from the time-averaged dynamics of its coupled resonant system. The model has been analyzed in both the weak and strong modulation regimes, rediscovering and providing new approximate expressions for the resulting relation. 
In particular, we derived an approximately linear relationship between the peak separation of the double-horned profile and the modulation amplitude, even in the nonlinear regime where the signal transduction is strongly distorted. 
\par
Furthermore, we have demonstrated that the proposed method can estimate the quality factor of a mechanical resonator and have shown the agreement between the model and the experimental results. The numerical simulations also qualitatively reproduce the observed optomechanical dynamics and support the interpretation of probe-induced dynamical backaction.
In particular, we have proved that, by scanning the probe across the optical resonance in both directions, the proposed method provides an in situ consistency check for such backaction while reducing its net influence when the average detuning over a scan is close to zero. Finally, the mechanical quality factor can be inferred without directly resolving or demodulating the mechanical oscillation.
\par
The present analytical treatment and experimental demonstration focus on the unresolved-sideband regime, $\Omega_m \ll \kappa$, where the optical cavity response follows the mechanical modulation adiabatically. Extending the ringdown method to the crossover and resolved-sideband regimes, $\Omega_m \gtrsim \kappa$, represents a natural direction for future work. 
More generally, the time-averaged response considered here is relevant to a broader class of resonant systems subject to coherent modulation, including examples in terahertz spectroscopy~\cite{Zhang2017-ve}, spin-resonance systems such as nitrogen-vacancy centers~\cite{arcizet_single_2011}, and modulated spectral-line measurements~\cite{Castignani2023-ys, Wiklind1997-nq}. The applicability of the present analytical description to such systems depends on the corresponding dynamical timescales and on the validity of an analogous time-averaged response approximation.\\

\textbf{Acknowledgments and Funding:} We acknowledge support from the Danish National Research Foundation (bigQ, DNRF0142).\\\\
\textbf{Data Availability Statement:}
The data that support the findings of this study are available from the corresponding author upon reasonable request.\\\\
\textbf{Author Contributions:}
D.A., D.T., and Z.A. conceptualized the study. D.T. and Z.A. developed the theoretical proof and methodology and performed the formal analysis. D.A. and Z.A. designed and conducted the experiments. Z.A. carried out the numerical analysis and acquired the data. D.T. and Z.A. curated the data. Z.A. wrote the original draft. D.T., D.A., A.H., and U.L.A. reviewed and edited the manuscript. A.H. and U.L.A. supervised the project. All authors have read and agreed to the published version of the manuscript.

\printbibliography{}

\appendix
\section{Mathematical methods}
\subsection{Analytical solution of the model}
\label{a:solution}
The integral in Equation~\ref{eq:integral_period_phi}, which shows the average behavior of a modulated Lorentzian peak, can be solved analytically using complex analysis. After substituting for a complex variable $z=e^{i\phi}$, we apply the residue theorem and solve the contour integral around a closed path $\abs{z} = 1$ (see Supplement~\ref{s:integral_solution} for the full derivation). The integral has four poles:
\begin{equation}
    m_1^\pm=\frac{-i\nu+1\pm\sqrt{(1-i\nu)^2+A^2}}{A},\quad m_2^\pm=\frac{-i\nu-1\pm\sqrt{(1+i\nu)^2+A^2}}{A}.
    \label{eq:a_poles}
\end{equation}
\par
It can be shown that only two of these ($m_1^-$ and $m_2^+$) lie within the unit circle for each value of $\nu$ and $A$, thus contributing to the integral, according to the definition of the residue theorem (see Supplement~\ref{s:moduli} for the proof). The complex solution of the integral is then given by
\begin{equation}
    \begin{split}
    \Lambda(\nu,A)&=\frac{1}{2\sqrt{(1-i\nu)^2+A^2}}+\frac{1}{2\sqrt{(1+i\nu)^2+A^2}}.  
    \end{split}
    \label{eq:a_integral_sqrt}
\end{equation}
From this identity, we can prove that the expression is real because it is the sum of two
complex-conjugate numbers (see Supplement~\ref{s:moduli}), and rewrite the solution as
\begin{equation}
\Lambda(\nu,A)=\frac{1}{2\zeta e^{-i\phi/2}}+\frac{1}{2\zeta e^{i\phi/2}}=\frac{1}{\zeta}\cos{\left(\phi/2\right)}=\sqrt{\frac{\sqrt{\eta_+\eta_-}+\eta_A}{2\eta_+\eta_-}},
\label{eq:a_integ}
\end{equation}
following the same definitions reported in the main text, where the modulus $\zeta$ and the argument $\phi$ are defined as $\zeta=\sqrt[\leftroot{0}\uproot{3}4]{\eta_+\eta_-}$ and  $\phi_\mp/2=\mp\frac{1}{2}\tan^{-1}{\left(2\nu/\eta_A\right)}+k\pi \text{ for }k=0,1$, respectively. 

\subsection{Approximation for the weak modulation limit}
\label{a:weak_approx}
In the limit $A\ll1$, the analytical expression can be approximated with Equation~\ref{eq:smallv_claim}. To prove this result, we can split the analysis for $\nu\rightarrow0$ and $\nu\rightarrow\infty$ and show that in both cases the two expressions converge.
In the limit $(\nu, A)\rightarrow(0,0)$, we can use Taylor's expansion of the model result and obtain
\begin{equation}\label{eq:weak_small_nu}
\Lambda(\nu,A)\simeq1-\nu^2-\frac{A^2}{2}+O(\nu^4+A^4+\nu^2A^2)\simeq \frac{1-\frac{A^2}{2}}{1+\nu^2}+O(\nu^4+A^4+\nu^2A^2),
\end{equation}
which proves the approximation.
\par
In the limit $(\abs{\nu}, A)\rightarrow(+\infty,0)$, we can use the same approach to rewrite the expression as
\begin{equation}\label{eq:weak_large_nu}
\Lambda(\nu,A)\simeq\frac{1}{\nu^2}-\frac{1-\frac{3}{2}A^2}{\nu^4}+O\left(\frac{1}{\nu^6}\right),
\end{equation}
so that the deviation from the approximated result is bounded and still grows negligibly for large $\abs{\nu}$:
\begin{equation}\label{eq:weak_small_nu_err}
\Lambda(\nu,A)-\frac{1-\frac{A^2}{2}}{1+\nu^2}\simeq\frac{A^2}{2\nu^2}+\frac{A^2}{\nu^4}+O\left(\frac{1}{\nu^6}\right)=O\left(\frac{A^2}{\nu^2}+\frac{1}{\nu^4}\right).
\end{equation}
\par
The complete derivation of these results is reported in Supplement~\ref{s:weak_regime}.

\subsection {Amplitude estimation in the strong modulation regime}
\label{a:strong_distance}

In the estimation of the modulation amplitude of a strong coherent drive, we investigate the derivative and stationary points of the analytical model in Equation~\ref{eq:real_solution}. The derivative is calculated as
\begin{equation}\label{eq:a_deriv}
  \frac{d\Lambda(\nu,A)}{d\nu}=  \frac{1}{\sqrt{2}}\frac{-\nu}{\left(\eta_+\eta_-\right)^{3/2}\sqrt{\sqrt{\eta_+\eta_-}+\eta_A}}\left[\eta_+\eta_-+\eta_\nu\left(2\eta_A+\sqrt{\eta_+\eta_-}\right)\right]
\end{equation}
(see Supplement~\ref{s:derivative} for the derivation), which can have one or three stationary points. For $\nu=0$, the derivative is always null, which was the maximum in the weak modulation regime and now becomes a relative minimum in the strong modulation regime. If $A>\sqrt{2}$, two new zeros appear for $\nu=\pm\nu_A$ by nullifying the expression between square brackets.
The solution to this polynomial equation can be rewritten as a third-order polynomial in $\nu^2$
\begin{equation}
    \nu^6+\left(1-\frac{3}{2}A^2\right)\nu^4-\left(1+3A^2\right)\nu^2-1-\frac{3}{2}A^2+\frac{1}{2}A^6=0,
\label{eq:v6_derivative}
\end{equation}
and it can be proven (see Supplement~\ref{s:derivative}) that it has three real solutions for $A>\sqrt{2}$: one is negative, so it is not allowed, one is positive and smaller than $A^2-1$, and the last is positive and larger.
By evaluating the sign of the derivative at $\nu^2=0$ and $\nu^2=A^2-1$ and invoking Descartes' rule of signs, we can recognize that the maxima correspond to the smaller solution, thus the statement $0<\nu_A<A$.
\par

Finally, we notice that when the constant term in Equation~\ref{eq:v6_derivative} ($-1-3/2A^2+1/2A^6$) is positive, these solutions appear, i.e., the resonance peak starts to split for $A>\sqrt{2}$.

\subsection{Approximation for the strong modulation regime}
\label{a:strong_approx}
The analytical approximation for the peaks' positions is derived from Equation~\ref{eq:derivative_solution}.
The polynomial can be reduced to a depressed cubic form $y^3+vy+w=0$ using the Tschirnhaus transformation $y=\eta_\nu+A^2/2-2/3$, obtaining
\begin{equation}
     v=-4A^2-\frac{1}{3}\left(\frac{3}{2}A^2-2\right)^2=-\frac{3}{4}\left(A^2+\frac{4}{3}\right)^2,\quad
    w=\frac{1}{4}\left(A^2-\frac{4}{3}\right)^3-\frac{8}{3}A^2.
\label{eq:a_depressed_cubic_coefficients}
\end{equation}
From this point, by using Cardano's formula, the solutions can be further simplified when written in trigonometric form because the discriminant $\Delta=(v/3)^3+(w/2)^2$ is negative and $v$ is a perfect square:
\begin{equation}
    y_k=2\sqrt{-\frac{v}{3}}\cos{\left[\frac{1}{3}\cos^{-1}\left(\frac{3w}{2v}\sqrt{-\frac{3}{v}}\right)-\frac{2k\pi}{3}\right]}=\left(A^2+\frac{4}{3}\right)\xi_k(\varphi),\quad k=0,1,2.
\label{eq:a_cubic_solutions_trigonometric}
\end{equation}
Here, the different values of k represent three possible solutions, as discussed in Appendix~\ref{a:strong_distance}.
\par
In Supplement~\ref{s:approx_peaks}, we derive this form and show that only one of these solutions satisfies the condition $0<\nu_A<A$, that is, the solution for $(k=1)$, thus obtaining
\begin{equation}
    \nu_A^2=-1+A^2-\frac{1}{3}\left(\frac{3}{2}A^2-2\right)+y_1\simeq A^2-\frac{2}{\sqrt{3}}A-\frac{1}{9}+O\left(\frac{1}{A}\right),
\label{eq:a_cubic_solution_1}
\end{equation}
which directly leads to the approximation reported in Equation~\ref{eq:approximation}.

\section{Numerical simulations of the optomechanical cavity response}
\label{a:numerical}

In addition to experimental validation, numerical results have also been analyzed. The simulation of our experimental setup follows the model of an optical cavity mode with frequency $\omega_0$ coupled to a mechanical resonator with frequency $\Omega_m$. In a frame rotating at the laser frequency $\omega(t)$, the system is described by the Hamiltonian
\begin{equation}
    H/\hbar = \Delta(t)\, \hat{a}^\dagger\hat{a} 
      + \Omega_m\, \hat{b}^\dagger \hat{b}
      + g_0\, \hat{a}^\dagger \hat{a} \bigl(\hat{b} + \hat{b}^\dagger\bigr)
      + i\sqrt{\kappa_\mathrm{ext}} \left(a_\mathrm{in}(t)\, \hat{a}^\dagger 
      - a_\mathrm{in}^*(t)\, \hat{a}\right),
\end{equation}
where $\hat{a}$ is the optical mode of the cavity and $\hat{b}$ is the mechanical mode of the system, $g_0$ is the single-photon optomechanical coupling strength, and $\kappa_{ext}$ is the coupling rate to the optical cavity. The time-dependent detuning is the difference between the scanning probe laser frequency and the optical cavity resonance, i.e., $\Delta(t) = \omega_0 - \omega(t)$.
The amplitude of the probing optical field $a_\mathrm{in}(t)$ is associated with the chosen input power $P_\mathrm{in}$ as
\begin{equation}
    a_\mathrm{in}(t) = \sqrt{\frac{P_\mathrm{in}}{\hbar \omega(t)}}e^{i\psi}.
\end{equation}
\par
Using the generalized form of the quantum Langevin equations in the Markov approximation \cite{bowenQuantumOptomechanics2016}, the equations of motion for the mean fields $a(t)$ and $b(t)$ can be derived using a semiclassical approach:

\begin{subequations}
\begin{align}
\frac{da}{dt}
&=
-\left[
\frac{\kappa}{2}
+i\Delta(t)
+i g_0\left(b+b^*\right)
\right]a
+\sqrt{\kappa_{\mathrm{ex}}}\,a_{\mathrm{in}}(t),
\\
\frac{db}{dt}
&=
-\left(
\frac{\Gamma_m}{2}
+i\Omega_m
\right)b
-i g_0|a|^2 .
\end{align}
\end{subequations}
where $\kappa$ is the optical decay rate and $\Gamma_m$ is the mechanical damping rate. Finally, the transmitted power of the cavity is derived from the input-output relations in the form $P_\mathrm{out}(t) = \hbar \omega(t)\kappa_\mathrm{ex}\abs{a(t)}^2$.

In this model, the mechanical mode is treated as a classical complex amplitude corresponding to a coherent state, so quantum and thermal noise are neglected. 
Thus, it can accurately describe the experimental results if the mechanical oscillator is initialized with a strong coherent state, e.g., $N_b=6 \times10^{9}$. The modulated resonance is probed with a certain optical power $P_\mathrm{in}$ by scanning the laser frequency $\omega(t)$ across the optical resonance with a triangular wave of period $T_w =200\mathrm{\mu s}$.
To mimic the slow measurement in experiments, we low-pass the transmitted power $P_\mathrm{out}(t)$ with a cutoff frequency equal to $f_s= 0.3(\Omega_{m}/2\pi)$ to extract the average dynamics.
\par
The time-domain simulations were performed in Python 3.12 using the SciPy RK45 integrator. 
Compared to the experimental parameters, we chose to increase the mechanical damping rate $\Gamma_m$ and the scanning speed $1/T_w$ to speed up the computation, without loss of generality. The other parameters are kept close to the experimental values. 
\par
\begin{figure}[h!]
    \centering
    \includegraphics[width=\textwidth]{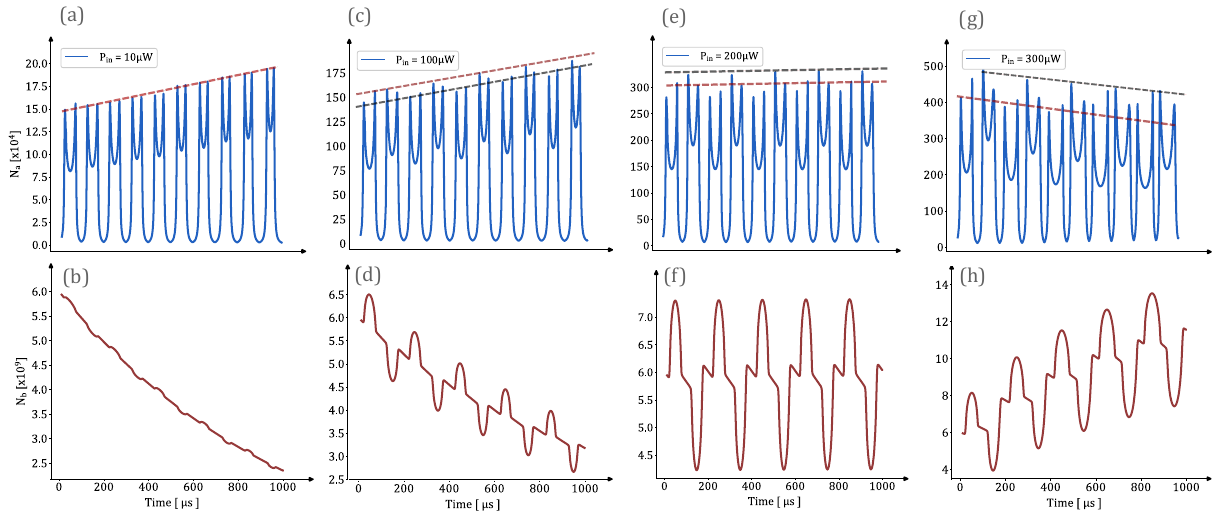}
    \caption{Time evolution of the optical resonance shape with changing power for fixed simulation parameters of $g_0/2\pi=1000\:\mathrm{Hz}$, $\kappa/2\pi=30 \:\mathrm{MHz}$, $\Gamma_m/2\pi=150\:\mathrm{Hz}$, $\omega_{0}/2\pi = 193.5 \:\mathrm{THz}$, scanning period $ T_{w}= 200\:\mathrm{\mu s}$, scanning range $\omega_0 - 300\:\mathrm{MHz} \to \omega_0 + 300\:\mathrm{MHz}$, and initialization of mechanical state $N_b = 6\cross10^9$. The plots show the evolution of the intracavity photon number $N_a$ (top) and phonon population $N_b$ (bottom) with different probing power $P_\mathrm{in}=10\:\mathrm{\mu W}$ (a-b), $100\:\mathrm{\mu W}$ (c-d), $200\:\mathrm{\mu W}$ (e-f), and $300\:\mathrm{\mu W}$ (g-h) respectively. The dashed lines represent the different trends for the red-to-blue and blue-to-red detuned scanning.
    }
    \label{fig:power_vs_chewing_gum_model}
\end{figure}
In Figure~\ref{fig:power_vs_chewing_gum_model}, we show the results of the numerical simulation of the number of intracavity photons $N_a$ and phonons $N_b$ over several consecutive scans of the optical resonance. 
When the probe power is small ($P_\mathrm{in}=10\:\mu{W}$), we verify that the mechanical population decays as expected because the probe's perturbations are negligible (see Figure \ref{fig:power_vs_chewing_gum_model}(b)). However, when we increase the power ($P_\mathrm{in}=100\:\mu{W}$), the direction of the probe's scan (from red to blue detuned or vice versa) becomes a factor, i.e., the mechanical oscillator is periodically heated and cooled by the scanning probe laser (see Figure \ref{fig:power_vs_chewing_gum_model}(c) and (d)). This effect becomes relevant when estimating the oscillator's quality factor because the effective decay rate is modified depending on the scanning side, as hinted by the dashed lines.  
As the probe power is increased, the optomechanical anti-damping becomes stronger and eventually balances the intrinsic mechanical damping. At this point,

\begin{equation}
    \Gamma_{\mathrm{eff}}
    =
    \Gamma_m+\Gamma_{\mathrm{opt}}
    =
    0.
\end{equation}
This condition marks the onset of parametric instability. Beyond this threshold, the mechanical mode no longer decays and instead enters a self-sustained oscillation regime, as shown in Figure~\ref{fig:power_vs_chewing_gum_model}(e).
\par
The threshold is also useful for estimating the strength of the optomechanical interaction. In the unresolved-sideband regime, the full expression for the optomechanical damping is given by,
\begin{equation}
\Gamma_{\mathrm{opt}}
=
g_0^2 n_{\mathrm{cav}}(\Delta_{\mathrm{eff}})
\left[
\frac{\kappa}
{(\kappa/2)^2+(\Delta_{\mathrm{eff}}-\Omega_m)^2}
-
\frac{\kappa}
{(\kappa/2)^2+(\Delta_{\mathrm{eff}}+\Omega_m)^2}
\right].
\end{equation}
where $\Delta_{\mathrm{eff}}=\Delta+g_0(b+b^*)$ is the effective cavity--laser detuning, including the optomechanically induced cavity-frequency shift, and
$n_{\mathrm{cav}}(\Delta_{\mathrm{eff}})
=
\kappa_{\mathrm{ex}}P_{\mathrm{in}}/
\left\{
\hbar\omega_L
\left[
(\kappa/2)^2+\Delta_{\mathrm{eff}}^2
\right]
\right\}$
is the detuning-dependent intracavity photon number.\\
Since $\Gamma_{\mathrm{opt}}=-\Gamma_m$ at the instability threshold, the measured threshold directly constrains the quantity $g_0^2 n_{\mathrm{cav}}$. If the intracavity photon number, cavity linewidth, and detuning are known, this relation can be used to estimate the single-photon optomechanical coupling rate $g_0$ \cite{Luan2014-oa}. Without an independent calibration of $n_{\mathrm{cav}}$, the measurement instead gives the pump-enhanced coupling $g_0\sqrt{n_{\mathrm{cav}}}$. An absolute measurement of single-photon coupling strength by tracing the onset of the Hopf bifurcation and limit cycles has also been shown in the literature in a similar way \cite{piergentili_absolute_2021}.

For sufficiently high input powers ($P_\mathrm{in}=\SI{300}{\micro\watt}$), as in Figure~\ref{fig:power_vs_chewing_gum_model}(g), the optomechanical gain exceeds the intrinsic mechanical dissipation ($\Gamma_\mathrm{opt}<-\Gamma_m$), resulting in negative effective damping ($\Gamma_\mathrm{eff}<0$). In this regime, the optical drive overcomes the mechanical losses and drives the resonator into a state of large-amplitude self-oscillations. This shows a potentially rich and complex interplay of nonlinear dynamics in a blue-detuned pump cavity optomechanical system in the unresolved-sideband regime.

\end{refsection}
\newpage
\begin{refsection}
\counterwithin{equation}{subsection}
\renewcommand{\theequation}{S\arabic{subsection}.\arabic{equation}}
\section[Appendix]{Supplementary information: complete derivation of the mathematical results of the paper}

\subsection[Appendix]{Analytical solution of the integral}\label{s:integral_solution}

The spectral response of a modulated Lorentzian peak averaged over a period $T$ is expressed as a time integral (see Equation~\ref{eq:integral_lorent} of the main text). After defining the normalized amplitude $A$ and the normalized detuning $\nu$, the integral is rewritten in terms of the modulation phase $\phi$ as
\begin{equation}
    \Lambda(\nu,A)=\frac{1}{2\pi}\int_{0}^{2\pi}\frac{d\phi}{1+\left(\nu+A\sin\phi\right)^2}.
    \label{eq:integral_phi}
\end{equation}
\vspace{\baselineskip}
\newline
This integral can be solved using the residue theorem after defining the complex variable $z=e^{i\phi}$ in a similar approach to other known integrals \cite{gradshteyn2014table}:
\begin{subequations}
    \begin{align}
    \Lambda(\nu,A)&=\frac{1}{2\pi}\ointop_{\abs{z}=1}\frac{1}{1+\left(\nu + A\frac{z-z^{-1}}{2i}\right)^2}\frac{dz}{iz}
    =\frac{1}{2\pi}\ointop_{\abs{z}=1}\frac{4iz}{-4z^2+\left(Az^2+2i\nu{z}-A\right)^2}dz
    \label{eq:integral_z}\\
    &=\frac{1}{2\pi}\ointop_{\abs{z}=1}\frac{4iz}{(Az^2+2(i\nu-1)z-A)(Az^2+2(i\nu+1)z-A)}dz\\
    &=\frac{1}{2\pi}\ointop_{\abs{z}=1}\frac{4iz}{A^2(z-m_1^+)(z-m_1^-)(z-m_2^+)(z-m_2^-)}dz\\
    &=-\frac{4}{A^2}\smashoperator{\sum_m}\Res(m)\quad\quad\mathrm{with}\ {m\in\left\{m_1^\pm,m_2^\pm\right\}\:\mathrm{s.t.}\:\abs{m}<1}
    \end{align}
    \label{eq:integral_res}
\end{subequations}
where $m_1^\pm$ and $m_2^\pm$ are the poles of the integrand, defined as
\begin{equation}
    m_1^\pm=\frac{-i\nu+1\pm\sqrt{(1-i\nu)^2+A^2}}{A},\quad m_2^\pm=\frac{-i\nu-1\pm\sqrt{(1+i\nu)^2+A^2}}{A}.
    \label{eq:poles}
\end{equation}
\newline\vspace{\baselineskip}
The residues are calculated as follows:
\begin{subequations}
    \begin{align}
    \Res(m_1^\pm)&=\lim_{z\rightarrow m_1^\pm}\frac{z}{(z-m_1^\mp)(z-m_2^+)(z-m_2^-)}=\frac{m_1^\pm}{(m_1^\pm-m_1^\mp)(m_1^\pm-m_2^+)(m_1^\pm-m_2^-)}\\
    &=\frac{A^2\left(-i\nu+1\pm\sqrt{(1-i\nu)^2+A^2}\right)}{\pm2\sqrt{(1-i\nu)^2+A^2}\left[\left(2\pm\sqrt{(1-i\nu)^2+A^2}\right)^2-(1+i\nu)^2-A^2\right]}\\
    &=\frac{A^2\left(-i\nu+1\pm\sqrt{(1-i\nu)^2+A^2}\right)}{\pm2\sqrt{(1-i\nu)^2+A^2}\left[4-4i\nu\pm4\sqrt{(1-i\nu)^2+A^2}\right]}\\
    &=\pm\frac{A^2}{8\sqrt{(1-i\nu)^2+A^2}}
    \end{align}
\end{subequations}
\begin{subequations}
    \begin{align}
    \Res(m_2^\pm)&=\lim_{z\rightarrow m_2^\pm}\frac{z}{(z-m_2^\mp)(z-m_1^+)(z-m_1^-)}=\frac{m_2^\pm}{(m_2^\pm-m_2^\mp)(m_2^\pm-m_1^+)(m_2^\pm-m_1^-)}\\
    &=\frac{A^2\left(-i\nu-1\pm\sqrt{(1+i\nu)^2+A^2}\right)}{\pm2\sqrt{(1+i\nu)^2+A^2}\left[\left(-2\pm\sqrt{(1+i\nu)^2+A^2}\right)^2-(1-i\nu)^2-A^2\right]}\\
    &=\frac{A^2\left(-i\nu-1\pm\sqrt{(1+i\nu)^2+A^2}\right)}{\pm2\sqrt{(1+i\nu)^2+A^2}\left[4+4i\nu\mp4\sqrt{(1+i\nu)^2+A^2}\right]}\\
    &=\mp\frac{A^2}{8\sqrt{(1+i\nu)^2+A^2}}
    \end{align}
\end{subequations}
\par
After verifying that the only two poles inside the unit circle are $m_1^-$ and $m_2^+$ (see Section~\ref{s:moduli}), we can write the final result for the integral as the following sum between two complex numbers:
\begin{equation}
    \begin{split}
    \Lambda(\nu,A)&=-\frac{4}{A^2}\left[\Res(m_1^-)+\Res(m_2^+)\right]=-\frac{4}{A^2}\left[-\frac{A^2}{8\sqrt{(1-i\nu)^2+A^2}}-\frac{A^2}{8\sqrt{(1+i\nu)^2+A^2}}\right]\\
    &=\frac{1}{2\sqrt{(1-i\nu)^2+A^2}}+\frac{1}{2\sqrt{(1+i\nu)^2+A^2}}    
    \end{split}
    \label{eq:integral_sqrt}
\end{equation}
\subsection[Appendix]{Roots moduli for the residue theorem and real-value solution}\label{s:moduli}
The study of the poles aims to determine which of them is within the contour of integration.
The first step of our analysis focuses on evaluating the moduli and angles of the square roots appearing in the four-pole expression (Equation~\ref{eq:poles}). After defining $\eta_A=1+A^2-\nu^2$ and $\eta_\pm=1+(A\pm\nu)^2$, we can use Proposition~\ref{eq:eta_pm_product} to rewrite the solution of square roots as
\begin{subequations}
    \begin{align}
    \zeta&=\abs{\sqrt{(1\mp i\nu)^2+A^2}}=\sqrt{\abs{
(1-i\nu)^2+A^2}}\\
&=\sqrt[\leftroot{0}\uproot{3}4]{(1-\nu^2+A^2)^2+4\nu^2}=\sqrt[\leftroot{0}\uproot{3}4]{\eta_A^2+4\nu^2}=\sqrt[\leftroot{0}\uproot{3}4]{\eta_+\eta_-}\\
    \phi_\mp/2&=\phase{\sqrt{(1\mp i\nu)^2+A^2}}=\frac{1}{2}\phase{(1\mp i\nu)^2+A^2}+k\pi\\
    &=\mp\frac{1}{2}\tan^{-1}{\left(\frac{2\nu}{\eta_A}\right)}+k\pi \quad\text{for }k=0,1
    \end{align}
\end{subequations}
so we can rewrite the poles, assuming $\phi_\mp/2\in\left(-\frac{\pi}{2},\frac{\pi}{2}\right)$ for $k=0$, as
\begin{subequations}
    \begin{align}
    m_1^\pm&=\frac{-i\nu+1\pm\zeta e^{i\phi_-}}{A},\quad\quad
    m_2^\pm=\frac{-i\nu-1\pm\zeta e^{i\phi_+}}{A}
    \end{align}
    \label{eq:poles_exponential}
\end{subequations}
\par
Using the following trigonometric identities (notice that the sign of $\sin(\phi_\mp/2)$ is determined by the sign of $\nu$, which is the imaginary part of the root)
\begin{subequations}
    \begin{align}
    \cos(\phi_\mp/2)&=\sqrt{\frac{1+\cos(\phi_\mp)}{2}}=\sqrt{\frac{1+\frac{\eta_A}{\sqrt{\eta_+\eta_-}}}{2}}=\sqrt{\frac{\sqrt{\eta_+\eta_-}+\eta_A}{2\sqrt{\eta_+\eta_-}}}\\
    \sin(\phi_\mp/2)&=\mp\sgn{(\nu)}\sqrt{\frac{1-\cos(\phi_\mp)}{2}}=\mp\sgn{(\nu)}\sqrt{\frac{\sqrt{\eta_+\eta_-}-\eta_A}{2\sqrt{\eta_+\eta_-}}}
    \end{align}
    \label{eq:trig_identities_cos_tan_over_2}
\end{subequations}
we can rewrite the poles in algebraic form as
\begin{subequations}
\begin{align}
m_1^\pm&=\frac{-i\nu+1}{A}\pm\frac{1}{\sqrt{2}A}\left(\sqrt{\sqrt{\eta_+\eta_-}+\eta_A}-i\sgn{(\nu)}\sqrt{\sqrt{\eta_+\eta_-}-\eta_A}\right)\\
m_2^\pm&=\frac{-i\nu-1}{A}\pm\frac{1}{\sqrt{2}A}\left(\sqrt{\sqrt{\eta_+\eta_-}+\eta_A}+i\sgn{(\nu)}\sqrt{\sqrt{\eta_+\eta_-}-\eta_A}\right)
\end{align}
\label{eq:poles_algebric}
\end{subequations}
\par
The four poles are related to each other by having the same imaginary part and the opposite real part ($m_1^\pm=-m_2^{\mp*}$). This is an expected property, since the starting integral was real and the residues must also be complex conjugate pairs. This allows us to consider only the two moduli of $m_1^+$ and $m_1^-$ to determine all the poles inside and outside the unit circle:
\begin{subequations}
\begin{align}
\abs{m_1^\pm}&=\frac{1}{A}\sqrt{\left(1\pm\frac{1}{\sqrt{2}}\sqrt{\sqrt{\eta_+\eta_-}+\eta_A}\right)^2+\left(-\nu\mp\frac{\sgn{(\nu)}}{\sqrt{2}}\sqrt{\sqrt{\eta_+\eta_-}-\eta_A}\right)^2}\\
&=\frac{1}{A}\sqrt{1+\nu^2+\sqrt{\eta_+\eta_-}\pm\sqrt{2}\left(\sqrt{\sqrt{\eta_+\eta_-}+\eta_A}+\abs{\nu}\sqrt{\sqrt{\eta_+\eta_-}-\eta_A}\right)}
\end{align}
\label{eq:poles_moduli}
\end{subequations}
\par
The result reported in Equation~\ref{eq:integral_sqrt} is verified if the conditions $\abs{m_1^+}>1$ and $\abs{m_1^-}<1$ are satisfied. This is equivalent to verifying that the following inequalities hold:
\begin{equation}
    \begin{cases}
        \eta_\nu+\sqrt{\eta_+\eta_-}+\sqrt{2}\left(\sqrt{\sqrt{\eta_+\eta_-}+\eta_A}+\abs{\nu}\sqrt{\sqrt{\eta_+\eta_-}-\eta_A}\right)>0\\
        \eta_\nu+\sqrt{\eta_+\eta_-}-\sqrt{2}\left(\sqrt{\sqrt{\eta_+\eta_-}+\eta_A}+\abs{\nu}\sqrt{\sqrt{\eta_+\eta_-}-\eta_A}\right)<0
    \end{cases}
\label{eq:inequalities_moduli}
\end{equation}
\par
The first inequality is trivially verified $\forall A$ using Proposition~\ref{eq:eta_pm_product} and removing the right part of the expression, since the square roots are always positive:
\begin{equation}
    \eta_\nu+\sqrt{\eta_+\eta_-}>1 \quad\Leftrightarrow\quad \eta_+\eta_->(\eta_\nu-1)^2 \quad\Leftrightarrow\quad 2\nu^2+2A^2 +1>0 \quad\forall \quad A,\nu
\label{eq:modulus_squared_gr_one}
\end{equation}
From this consideration, we conclude that the average of the squared moduli $\left(\abs{m_1^+}^2+\abs{m_1^-}^2\right)/2$ is always greater than one. 
We can also calculate that the average of the moduli themselves is always greater than one. In fact, since the product of the roots is $m_1^+m_1^-=-1$, Equation~\ref{eq:poles_moduli} gives
\begin{equation}
\begin{aligned}
    \frac{\abs{m_1^+}+\abs{m_1^-}}{2}
    &=\frac{1}{2}\sqrt{\abs{m_1^+}^2+\abs{m_1^-}^2+2\abs{m_1^+}\abs{m_1^-}}
    =\frac{1}{2}\sqrt{\frac{2\left(1+\nu^2+\sqrt{\eta_+\eta_-}\right)}{A^2}+2}\\
    &=\sqrt{\frac{1+\nu^2+A^2+\sqrt{\eta_+\eta_-}}{2A^2}}
    =\sqrt{1+\frac{\eta_\nu+\sqrt{\eta_+\eta_-}}{2A^2}}>1 .
\end{aligned}
\end{equation}
\par
For the second inequality, we can square the two terms and rewrite the condition using Proposition~\ref{eq:eta_pm_product}, \ref{eq:eta_A_v_squared_diff}, and \ref{eq:eta_A_v_sum_diff}, and the definition $\eta_\nu=1+\nu^2-A^2$. We can prove it $\forall A^2>0$ with the following chain of inequalities:
\begin{subequations}
    \begin{align}    
        &\begin{cases}
            \left(\eta_\nu+\sqrt{\eta_+\eta_-}\right)^2<2\left(\sqrt{\sqrt{\eta_+\eta_-}+\eta_A}+\abs{\nu}\sqrt{\sqrt{\eta_+\eta_-}-\eta_A}\right)^2\\
            \eta_\nu+\sqrt{\eta_+\eta_-}>0 \quad\text{since if $<0$ the original inequality is always verified}
        \end{cases}\\
        &\Leftrightarrow\begin{cases}
            \eta_\nu^2+\eta_+\eta_-+2\eta_\nu\sqrt{\eta_+\eta_-}<2(1+\nu^2)\sqrt{\eta_+\eta_-}+2(1-\nu^2)\eta_A+4\abs{\nu}\sqrt{\eta_+\eta_--\eta_A^2}\\
            \eta_A^2>0
        \end{cases}\\
        &\Leftrightarrow\ \eta_\nu^2+\eta_A^2+4\nu^2+2\eta_\nu\sqrt{\eta_+\eta_-}<2(\eta_\nu+A^2)\sqrt{\eta_+\eta_-}+2(\eta_A-A^2)\eta_A+8\nu^2\\
        &\Leftrightarrow\ \eta_\nu^2-\eta_A^2-4\nu^2+2A^2\eta_A<2A^2\sqrt{\eta_+\eta_-}\\
        &\Leftrightarrow\ 2A^2(-2+\eta_A)<2A^2\sqrt{\eta_+\eta_-}\\
        &\Leftrightarrow\begin{cases}
            A^2>0\\
            -\eta_\nu<\sqrt{\eta_+\eta_-}
        \end{cases} \quad\Leftrightarrow\quad
        \eta_\nu^2<\eta_\nu^2+4A^2 \quad\Leftrightarrow\quad A^2>0
    \end{align}
    \label{eq:modulus_less_one}
\end{subequations}
\vspace{\baselineskip}
\par
This analysis can also be used to rewrite the integral solution $\Lambda(\nu,A)$ as a real expression. In fact, Equation~\ref{eq:integral_sqrt} can be rewritten as
\begin{equation}
\Lambda(\nu,A)=\frac{1}{2\zeta e^{-i\phi/2}}+\frac{1}{2\zeta e^{i\phi/2}}=\frac{1}{\zeta}\cos{\left(\phi/2\right)}=\sqrt{\frac{\sqrt{\eta_+\eta_-}+\eta_A}{2\eta_+\eta_-}}
\label{eq:integral_real}
\end{equation}
\par
As a final remark, we can discuss that for the limit $A\rightarrow0$, the smaller poles tend to $0$, so that ${\eta_+\eta_-}\rightarrow(1+\nu^2)^2$ and $\eta_A\rightarrow1-\nu^2$. We can thus rediscover the integral solution as a simple Lorentzian function:
\begin{equation}
\Lambda(\nu,0)=\lim_{A\rightarrow0}\Lambda(\nu,A)=\sqrt{\frac{(1+\nu^2)+(1-\nu^2)}{2(1+\nu^2)^2}}=\frac{1}{1+\nu^2}
\label{eq:integral_lorentzian}
\end{equation}

\subsection[Appendix]{Approximation in the weak modulation limit}
\label{s:weak_regime}
In addition to showing the limit for $A\rightarrow0$, we can derive an approximate result valid in the weak-drive regime ($A\ll1$). In the following, we will prove that the integral solution in Equation~\ref{eq:integral_real} is approximated by a Lorentzian peak $\mathcal{L}_1(\nu,A)$ reduced by a factor proportional to the squared modulation amplitude:
\begin{equation}\label{eq:L1_approximation}
\Lambda(\nu,A) \approx \mathcal{L}_1(\nu,A):=\left(1-\frac{A^2}{2}\right)\frac{1}{1+\nu^2}.
\end{equation}
\par
The proof of this approximation is performed by showing the convergence between the two expressions for the two limits $\nu\rightarrow0$ and $\nu\rightarrow+\infty$.
\par
In the regime $\nu\ll1$, we can approximate the term $\eta_+\eta_-\sim1+2(A^2+\nu^2)+\mathcal{O}(\nu^4+A^4+A^2\nu^2)$ using Taylor's expansion, and, with the same approach, we can rewrite $\Lambda(\nu,A)$ as
\begin{subequations}
    \begin{align}
    \Lambda(\nu,A)&\sim\sqrt{\frac{\sqrt{1+2(A^2+\nu^2)+\mathcal{O}(\nu^4+A^4+A^2\nu^2)}+1+A^2-\nu^2}{2+4(A^2+\nu^2)+\mathcal{O}(\nu^4+A^4+A^2\nu^2)}}\\
    &\sim\sqrt{\frac{1+A^2+\mathcal{O}(\nu^4+A^4+A^2\nu^2)}{1+2A^2+2\nu^2+\mathcal{O}(\nu^4+A^4+A^2\nu^2)}}\\
    &\sim\sqrt{\left(1+A^2+\mathcal{O}(\nu^4+A^4+A^2\nu^2)\right)\left(1-2A^2-2\nu^2+\mathcal{O}(\nu^4+A^4+A^2\nu^2)\right)}\\
    &\sim\sqrt{1-A^2-2\nu^2+\mathcal{O}(\nu^4+A^4+A^2\nu^2)}\\
    &\sim1-\frac{A^2}{2}-\nu^2+\mathcal{O}(\nu^4+A^4+A^2\nu^2)\sim\frac{1-\frac{A^2}{2}}{1+\nu^2}+\mathcal{O}(A^4+A^2\nu^2)
    \end{align}
    \label{eq:L1_smallnu}
\end{subequations}
thus proving the equality.
\par
In the regime $\nu\gg1$, we can proceed with a similar approach to show an asymptotic form for both the integral solution
\begin{subequations}
    \begin{align}
    \Lambda(\nu,A)&\sim\sqrt{\frac{\nu^2\sqrt{1+\frac{2(1-A^2)}{\nu^2}+\frac{(1+A^2)^2}{\nu^4}}+1+A^2-\nu^2}{2\nu^4\left(1+\frac{2(1-A^2)}{\nu^2}+\mathcal{O}\left(\frac{1}{\nu^4}\right)\right)}}\\
    &\sim\frac{1}{\nu^2}\sqrt{\frac{1+\frac{A^2}{\nu^2}+\mathcal{O}\left(\frac{1}{\nu^4}\right)}{1+\frac{2(1-A^2)}{\nu^2}+\mathcal{O}\left(\frac{1}{\nu^4}\right)}}\\
    &\sim\frac{1}{\nu^2}\sqrt{\left(1+\frac{A^2}{\nu^2}+\mathcal{O}\left(\frac{1}{\nu^4}\right)\right)\left(1-\frac{2(1-A^2)}{\nu^2}+\mathcal{O}\left(\frac{1}{\nu^4}\right)\right)}\\
    &\sim\frac{1}{\nu^2}\sqrt{1-\frac{2-3A^2}{\nu^2}+\mathcal{O}\left(\frac{1}{\nu^4}\right)}\sim\frac{1}{\nu^2}-\frac{1-\frac{3A^2}{2}}{\nu^4}+\mathcal{O}\left(\frac{1}{\nu^6}\right)
    \end{align}
\end{subequations}
and the approximation in Equation~\ref{eq:L1_approximation}
\begin{equation}
    \mathcal{L}_1(\nu,A)\sim\frac{1-\frac{A^2}{2}}{\nu^2}-\frac{1-\frac{A^2}{2}}{\nu^4}+\mathcal{O}\left(\frac{1}{\nu^6}\right).
\end{equation}
We can then bound the difference with a term which decreases at the same rate as the tails of the Lorentzian
\begin{equation}
    \Lambda(\nu,A)-\mathcal{L}_1(\nu,A)\sim\frac{A^2}{2\nu^2}+\frac{A^2}{\nu^4}+\mathcal{O}\left(\frac{1}{\nu^6}\right)\sim\mathcal{O}\left(\frac{A^2}{\nu^2} + \frac{1}{\nu^4}\right).
\end{equation}
In the considered context, this difference in the tails is negligible because most of the information is contained in the peak of the response.
Finally, we can write a bound for the difference between the approximated and analytical results in both regimes as
\begin{equation}
    \Lambda(\nu,A)-\mathcal{L}_1(\nu,A)\sim\mathcal{O}\left(A^4+\min\left(A^2\nu^2 + \nu^4, \frac{A^2}{\nu^2} + \frac{1}{\nu^4}\right)\right).
\end{equation}
\subsection[Appendix]{Derivative of the spectral response and stationary points}\label{s:derivative}
An important property of the spectral response is that for $A>\sqrt{2}$, the spectrum has a minimum at $\nu=0$ and two maxima at $\nu=\pm\nu_A$. This can be verified by calculating the derivative of the integral solution (Equation~\ref{eq:integral_real}) with respect to $\nu$ using Proposition~\ref{eq:part_derv}:
\begin{subequations}
    \begin{align}
    \frac{d\Lambda(\nu,A)}{d\nu}&=\frac{1}{\sqrt{2}}\left[\frac{\partial}{\partial\nu}\left(\frac{1}{\sqrt{\eta_+\eta_-}}\right)\sqrt{\sqrt{\eta_+\eta_-}+\eta_A}+\frac{1}{\sqrt{\eta_+\eta_-}}\frac{\partial}{\partial\nu}\left(\sqrt{\sqrt{\eta_+\eta_-}+\eta_A}\right)\right]\\
    &=\frac{1}{\sqrt{2}}\left[-\frac{4\nu\eta_\nu}{2\left(\eta_+\eta_-\right)^{3/2}}\sqrt{\sqrt{\eta_+\eta_-}+\eta_A}+\frac{1}{\sqrt{\eta_+\eta_-}}\frac{1}{2\sqrt{\sqrt{\eta_+\eta_-}+\eta_A}}\left(\frac{4\nu\eta_\nu}{2\sqrt{\eta_+\eta_-}}-2\nu\right)\right]\\
    &=\frac{1}{\sqrt{2}}\frac{-2\nu\eta_\nu(\sqrt{\eta_+\eta_-}+\eta_A)+\nu\eta_\nu\sqrt{\eta_+\eta_-}-\nu\eta_+\eta_-}{\left(\eta_+\eta_-\right)^{3/2}\sqrt{\sqrt{\eta_+\eta_-}+\eta_A}}\\
    &=\frac{1}{\sqrt{2}}\frac{-\nu}{\left(\eta_+\eta_-\right)^{3/2}\sqrt{\sqrt{\eta_+\eta_-}+\eta_A}}\left[\eta_+\eta_-+\eta_\nu\left(2\eta_A+\sqrt{\eta_+\eta_-}\right)\right]
    \end{align}
    \label{eq:derivative_integral}
\end{subequations}
\par
From this expression, we can easily observe the stationary point for $\nu=0$. The other stationary points can be found by setting the factor in the brackets to zero.
\par
We start with noticing that the stationary points $\pm\nu_A$ must satisfy $\abs{\nu_A}<A$ when $A^2>2$ because the derivative must cross $0$ between $\nu=0$ and $\abs{\nu}=\sqrt{A^2-1}<A$ since
\begin{equation}
    \begin{cases}
        \lim\limits_{\abs{\nu}\to 0^+}\eta_+\eta_-+\eta_\nu\left(2\eta_A+\sqrt{\eta_+\eta_-}\right)=-2(A^4-A^2-2)=-2(A^2-2)(A^2+1)=l_-<0,\ \forall A^2>2\\
        \lim\limits_{\abs{\nu}\to \sqrt{A^2-1}^-}\eta_+\eta_-+\eta_\nu\left(2\eta_A+\sqrt{\eta_+\eta_-}\right)=4A^2=l_+>0
    \end{cases}
\end{equation}
This also shows that $0$ is a minimum of the spectrum because the total derivative changes sign from negative to positive when moving from $\nu<0$ to $\nu>0$, while $\pm\nu_A$ are two maxima, if they exist.
Now, we can try to solve the derivative to find the stationary point $\nu_A$ using the fact that $\eta_\nu<0$ and $\eta_A>0$ for $0<\nu<\sqrt{A^2-1}$. Together with Proposition~\ref{eq:eta_pm_product} and \ref{eq:eta_pm_eta_A_v}, we can rewrite the inequality as
{\allowdisplaybreaks
\begin{subequations}
    \begin{align}    
    &\eta_+\eta_-+\eta_\nu\left(2\eta_A+\sqrt{\eta_+\eta_-}\right)=\eta_+\eta_-+\eta_\nu\left(4-2\eta_\nu+\sqrt{\eta_+\eta_-}\right)>0\\
    &\Leftrightarrow\quad
    \begin{cases}
        \left(\eta_+\eta_-+2\eta_\nu\eta_A\right)^2>\eta_\nu^2\eta_+\eta_-\\
        \eta_+\eta_-+2\eta_\nu\eta_A>0
    \end{cases}\\
    &\Leftrightarrow\quad
    \begin{cases}
        \eta_+\eta_-\left(\eta_+\eta_-+4\eta_\nu\eta_A-\eta_\nu^2\right)+4\eta_\nu^2\eta_A^2>0\\
        \eta_A(\eta_A+2\eta_\nu)+4\nu^2=\eta_A(4-\eta_A)+4\nu^2=4(1+A^2)-\eta_A^2>0
    \end{cases}\\
    &\Leftrightarrow\quad
    \begin{cases}
        \eta_+\eta_-\left(\eta_\nu\eta_A+A^2\right)+\eta_\nu^2\eta_A^2=2(1+\nu^2+A^2)\eta_\nu\eta_A+A^2\eta_+\eta_->0\\
        \nu^4-2\nu^2(1+A^2)+(A^2-3)(1+A^2)<0
    \end{cases}\\
    &\Leftrightarrow\quad
    \begin{cases}
        \eta_\nu\eta_A(2+2\nu^2+A^2)+2A^2(1+\nu^2+A^2)>0\\
        \Delta^2/4=(1+A^2)^2-(A^2-3)(1+A^2)=4(1+A^2)>0 \quad \forall A
    \end{cases}\\
    &\Leftrightarrow\quad
    \begin{cases}
        -2\nu^6+(-2+3A^2)\nu^4+(2+6A^2)\nu^2+(2+3A^2-A^6)>0\\
        \eta_A<2\sqrt{1+A^2} \;\Leftrightarrow\; \nu^2>1+A^2-2\sqrt{1+A^2} 
    \end{cases}\\
    &\Leftrightarrow\quad\nu^6+(1-3/2A^2)\nu^4-(1+3A^2)\nu^2-1-3/2A^2+1/2A^6<0
    \end{align}
\label{eq:derivative_nu}
\end{subequations}
}
In the last step, the second condition is absorbed by the first one within $0<\nu^2<A^2-1$. The polynomial in the last line is positive at $\nu^2=0$ and negative at $\nu^2=A^2-1$, so it has a single root $\nu_A^2$ in this interval and is negative only for $\nu_A^2<\nu^2<A^2-1$.
\par
The final equation can be solved analytically or numerically, producing three real solutions for $\nu_A^2$. However, only one respects the condition $0<\abs{\nu_A}<A$, while the root larger than $A^2-1$ belongs to the branch $\eta_+\eta_-+2\eta_\nu\eta_A=+\eta_\nu\sqrt{\eta_+\eta_-}$ introduced by squaring. In Section~\ref{s:approx_peaks}, we will show a way to approximate these solutions and discuss their behavior in detail.
\par
Finally, we can verify for which intervals of $A$ the two maxima exist by analysing the associated third-order equation in $\nu^2$. We first notice that the polynomial never has three positive solutions, and it has two positive solutions only if the last term, i.e., the intercept of the polynomial, is positive $\left(-1-3/2A^2+1/2A^6>0\right)$, which is satisfied for $A>\sqrt{2}$.
In fact, the limit for $A\to\sqrt{2}^+$ gives as solutions $\nu_A=0$ with multiplicity two and $\abs{\nu_A}=\sqrt{1+\sqrt{8}}>A=\sqrt{2}$ (which cannot be accepted by the previous considerations). The result allows us to argue, for continuity of the derivative, that the first solution becomes negative for $A<\sqrt{2}$, while the second remains positive for $A>0$, and it is the one we are interested in.
\par
The analysis proves that the Lorentzian peak splits for $A>\sqrt{2}$, and the separation between the two new maxima grows monotonically with $A$.
\subsection[Appendix]{Approximate solution for the peaks' separation}\label{s:approx_peaks}
To derive an analytical approximation for the positions of the two peaks, we start from Equation~\ref{eq:derivative_integral}, but instead of deriving an expression for the stationary points $\nu_A$, we rewrite the equation as a polynomial in the variable $\eta_\nu$. The advantage of this approach will become apparent in the analysis of the polynomial's roots.
Using Proposition~\ref{eq:eta_pm_product} and \ref{eq:eta_A_v_sum_diff}, we can write
\begin{subequations}
    \begin{align}
    &\eta_+\eta_-+\eta_\nu\left(2\eta_A+\sqrt{\eta_+\eta_-}\right)=\eta_+\eta_-+\eta_\nu\left(4-2\eta_\nu+\sqrt{\eta_+\eta_-}\right)>0\\
    &\Leftrightarrow\quad
    \begin{cases}
        \left(-\eta_\nu^2+4A^2+4\eta_\nu\right)^2>\eta_\nu^4+4A^2\eta_\nu^2\\
        \eta_+\eta_-+2\eta_\nu\eta_A>0
    \end{cases}\\
    &\Leftrightarrow\quad \eta_\nu^4+16(A^2+\eta_\nu)^2-8\eta_\nu^3-8A^2\eta_\nu^2>\eta_\nu^4+4A^2\eta_\nu^2\\
    &\Leftrightarrow\quad -8\eta_\nu^3+(16-12A^2)\eta_\nu^2+32A^2\eta_\nu+16A^4>0\\
    &\Leftrightarrow\quad \eta_\nu^3+(3/2A^2-2)\eta_\nu^2-4A^2\eta_\nu-2A^4<0
    \end{align}
\label{eq:derivative_eta_nu}
\end{subequations}
\par
This polynomial can be reduced to a depressed cubic equation by using the Tschirnhaus Transformation $y=\eta_\nu+p/3$ into the form $y^3+vy+w=0$ with the coefficients defined as
\begin{subequations}
    \begin{align}
    p&=\frac{3}{2}A^2-2,\quad\quad q=-4A^2,\quad\quad r=-2A^4\\
    v&=q-\frac{p^2}{3}=-4A^2-\frac{1}{3}\left(\frac{3}{2}A^2-2\right)^2=-\frac{3}{4}\left(A^2+\frac{4}{3}\right)^2\\
    w&=r+\frac{2p^3}{27}-\frac{pq}{3}=-2A^4+\frac{1}{4}\left(A^2-\frac{4}{3}\right)^3+2A^2\left(A^2-\frac{4}{3}\right)=\frac{1}{4}\left(A^2-\frac{4}{3}\right)^3-\frac{8}{3}A^2
    \end{align}
\label{eq:depressed_cubic_coefficients}
\end{subequations}
\par
In general, the roots of the depressed cubic equation can be found using Cardano's Method; however, in this case, the solution can be further simplified using the trigonometric solutions because $v$ is a perfect square. To apply this formula, we need to verify that the equation has three real roots, which is true if the discriminant $\Delta=(v/3)^3+(w/2)^2$ is negative:
{\allowdisplaybreaks
\begin{subequations}
    \begin{align}
    \Delta&=\left(-\frac{1}{4}\left(A^2+\frac{4}{3}\right)^2\right)^3+\left(\frac{1}{8}\left(A^2-\frac{4}{3}\right)^3-\frac{4}{3}A^2\right)^2<0\\
    &\Leftrightarrow\quad \left(A^2+\frac{4}{3}\right)^6>\left(\left(A^2-\frac{4}{3}\right)^3-\frac{32}{3}A^2\right)^2\\
    &\Leftrightarrow\quad
    \begin{cases}
        \begin{cases}
        \left(A^2-\frac{4}{3}\right)^3-\frac{32}{3}A^2<0\\
        \left(A^2+\frac{4}{3}\right)^3+\left(A^2-\frac{4}{3}\right)^3-\frac{32}{3}A^2>0
        \end{cases}\\
        \begin{cases}
        \left(A^2-\frac{4}{3}\right)^3-\frac{32}{3}A^2>0\\
        \left(A^2+\frac{4}{3}\right)^3-\left(A^2-\frac{4}{3}\right)^3+\frac{32}{3}A^2>0
        \end{cases}
    \end{cases}\\
    &\Leftrightarrow\quad
    \begin{cases}
        \begin{cases}
        \left(A^2-\frac{4}{3}\right)^3-\frac{32}{3}A^2<0\\
        2A^2\left(A^4+\frac{16}{3}-\frac{16}{3}\right)>0 \quad\Leftrightarrow\quad A^2>0
        \end{cases}\\
        \begin{cases}
        \left(A^2-\frac{4}{3}\right)^3-\frac{32}{3}A^2>0\\
        \frac{8}{3}\left(3A^4+\frac{16}{9}+4A^2\right)>0 \quad\Leftrightarrow\quad \Delta^2/4=4-\frac{16}{3}<0 \quad\Leftrightarrow\quad \forall A^2
        \end{cases}
    \end{cases}
    \end{align}
\label{eq:discriminant_depressed_cubic}
\end{subequations}
}
We can conclude that this is verified for $A>0$ even without solving the top inequality because $A=0$ is included in the first interval, and the bottom inequalities are true for every other value of $A$.
\par
The three trigonometric solutions $y_k$ are written as
\begin{subequations}
    \begin{align}
    y_k&=2\sqrt{-\frac{v}{3}}\cos{\left[\frac{1}{3}\cos^{-1}\left(\frac{3w}{2v}\sqrt{-\frac{3}{v}}\right)-\frac{2k\pi}{3}\right]},\quad k=0,1,2\\
    &=\left(A^2+\frac{4}{3}\right)\cos{\left[\frac{1}{3}\cos^{-1}\left(\frac{32A^2/3-(A^2-4/3)^3}{\left(A^2+4/3\right)^3}\right)-\frac{2k\pi}{3}\right]},\quad k=0,1,2\\
    &=\left(A^2+\frac{4}{3}\right)\xi_k(\varphi).
    \end{align}
\label{eq:cubic_solutions_trigonometric}
\end{subequations}
\par
To further the analysis, we consider the solutions in the limit $A\to\infty$. The first solution $(k=0)$ cannot be accepted because $\nu_A\sim A+1/\sqrt{3}>A$ which contradicts the previously found conditions about the maxima:
\begin{subequations}
    \begin{align}
    &\varphi\sim -1+\frac{8}{A^2}-\frac{64}{3A^4}+\frac{1280}{27A^6}+\mathcal{O}\left(\frac{1}{A^8}\right)\\
    &\xi_0(\varphi)=\cos(\frac{1}{3}\cos^{-1}(\varphi))\sim\frac{1}{2}+\frac{2}{\sqrt{3}A}-\frac{4}{9A^2}-\frac{52}{27\sqrt{3}A^3}+\mathcal{O}\left(\frac{1}{A^4}\right)\\
    &y_0=\left(A^2+\frac{4}{3}\right)\xi_0(\varphi)\sim\frac{A^2}{2}+\frac{2}{\sqrt{3}}A+\frac{2}{9}+\frac{20}{27\sqrt{3}}A^{-1}+\mathcal{O}\left(\frac{1}{A^2}\right)\\
    &\nu_A^2=-1+A^2-\frac{1}{3}\left(\frac{3}{2}A^2-2\right)+y_0\sim A^2+\frac{2}{\sqrt{3}}A-\frac{1}{9}+\mathcal{O}\left(\frac{1}{A}\right)\\
    &\nu_A\sim\sqrt{A^2+\frac{2}{\sqrt{3}}A-\frac{1}{9}+\mathcal{O}\left(\frac{1}{A}\right)}\sim A+\frac{1}{\sqrt{3}}+\mathcal{O}\left(\frac{1}{A}\right)>A.
    \end{align}
\label{eq:cubic_solution_0}
\end{subequations}
The second solution $(k=1)$ is the one we are looking for since $0<\nu_A<A$:
\begin{subequations}
    \begin{align}
    &\xi_1(\varphi)=\cos(\frac{1}{3}\cos^{-1}(\varphi)-\frac{2\pi}{3})=-\sin(\frac{1}{3}\sin^{-1}(\varphi))\\
    &\hspace{12em}\sim\frac{1}{2}-\frac{2}{\sqrt{3}A}-\frac{4}{9A^2}+\frac{52}{27\sqrt{3}A^3}+\mathcal{O}\left(\frac{1}{A^4}\right)\\
    &y_1=\left(A^2+\frac{4}{3}\right)\xi_1(\varphi)\sim\frac{A^2}{2}-\frac{2}{\sqrt{3}}A+\frac{2}{9}-\frac{20}{27\sqrt{3}}A^{-1}+\mathcal{O}\left(\frac{1}{A^2}\right)\\
    &\nu_A^2=-1+A^2-\frac{1}{3}\left(\frac{3}{2}A^2-2\right)+y_1\sim A^2-\frac{2}{\sqrt{3}}A-\frac{1}{9}+\mathcal{O}\left(\frac{1}{A}\right)\\
    &\nu_A\sim\sqrt{A^2-\frac{2}{\sqrt{3}}A-\frac{1}{9}+\mathcal{O}\left(\frac{1}{A}\right)}\sim A-\frac{1}{\sqrt{3}}+\mathcal{O}\left(\frac{1}{A}\right)<A.
    \end{align}
\label{eq:cubic_solution_1}
\end{subequations}
Finally, the third solution $(k=2)$ is always negative and cannot be accepted:
\begin{subequations}
    \begin{align}
    &\xi_2(\varphi)=\cos(\frac{1}{3}\cos^{-1}(\varphi)-\frac{4\pi}{3})\sim-1+\frac{8}{9A^2}-\frac{320}{243A^4}+\mathcal{O}\left(\frac{1}{A^6}\right)\\
    &y_2=\left(A^2+\frac{4}{3}\right)\xi_2(\varphi)\sim-A^2-\frac{4}{9}-\frac{32}{243}A^{-2}+\mathcal{O}\left(\frac{1}{A^4}\right)\\
    &\nu_A^2\sim-\frac{1}{2}A^2-\frac{7}{9}+\mathcal{O}\left(\frac{1}{A^2}\right)<0.
    \end{align}
\label{eq:cubic_solution_2}
\end{subequations}
\par
The only solution that satisfies the condition $0<\nu_A<A$ is the second one, which has an asymptote as $\nu_A=A-{1}/{\sqrt3}$. This means that the two maxima of the spectrum are separated by a distance $\Delta\nu=2\nu_A\sim 2(A-1/\sqrt{3})$, which grows linearly with $A$. 
Furthermore, we can numerically verify that the approximation is already very good for small values of $A$, completely satisfying the requirements for data analysis and experimental measurements.
\par
As a final consideration, we can compare the solution domain with the condition expressed in Section~\ref{s:derivative} for the exact solution, i.e., $0<\nu_A<\sqrt{A^2-1}$. Since the two inequalities are always verified within the domain, we can conclude that our approximation is valid for each value of $A>\sqrt{2}$, where an exact solution exists.
\subsection[Appendix]{Properties and useful identities}\label{s:properties}
In the derivation of the integral solution, we used several quantities and identities that allow for simplifying the analysis.
\par
With the definition of the quantities $\eta_\pm$, $\eta_A$, and $\eta_\nu$ as
\begin{equation}
\eta_\pm:=1+(A\pm\nu)^2,\quad\quad
\eta_A:=1+A^2-\nu^2,\quad\quad
\eta_\nu:=1+\nu^2-A^2,
\label{eq:eta_definitions}
\end{equation}
the following propositions can be proven:
\begin{proposition}
Given the product $\eta_+\eta_-$, it can be rewritten in terms of $\eta_A$ and $\eta_\nu$ as
\begin{equation}
    \eta_+\eta_-=\eta_A^2+4\nu^2=\eta_\nu^2+4A^2.
\label{eq:eta_pm_product}
\end{equation}
\textbf{Proof:}
\begin{subequations}\nonumber
    \begin{align}
    \eta_+\eta_-&=(1+(A+\nu)^2)(1+(A-\nu)^2)=1+(A+\nu)^2(A-\nu)^2+(A+\nu)^2+(A-\nu)^2\\
    &=1+(A^2-\nu^2)^2+2(A^2+\nu^2)=1+(A^2-\nu^2)^2+2(A^2-\nu^2)+4\nu^2\\
    &=(1+A^2-\nu^2)^2+4\nu^2=\eta_A^2+4\nu^2\\
    &=1+(A^2-\nu^2)^2-2(A^2-\nu^2)+4A^2=(1+\nu^2-A^2)^2+4A^2=\eta_\nu^2+4A^2
    \end{align}
\end{subequations}
\end{proposition}
\begin{proposition}
Given $\eta_A$ and $\eta_\nu$, the following identities are valid:
\begin{equation}
    \eta_A=2-\eta_\nu,\quad\quad
    \eta_A+\eta_\nu=2,\quad\quad
    \eta_A-\eta_\nu=2(A^2-\nu^2). 
\label{eq:eta_A_v_sum_diff}
\end{equation}
\textbf{Proof:}
\begin{subequations}\nonumber
\begin{align}
    &\eta_A=1+A^2-\nu^2=2-(1-A^2+\nu^2)=2-\eta_\nu\\
    &\eta_A+\eta_\nu=(1+A^2-\nu^2)+(1+\nu^2-A^2)=2\\
    &\eta_A-\eta_\nu=(1+A^2-\nu^2)-(1+\nu^2-A^2)=2(A^2-\nu^2)
\end{align}
\end{subequations}
\end{proposition}
\begin{proposition}
Given $\eta_A^2$ and $\eta_\nu^2$, their difference can be expressed as
\begin{equation}
    \eta_A^2-\eta_\nu^2=4(A^2-\nu^2)=2(\eta_A-\eta_\nu)=4\eta_A-4=4-4\eta_\nu.
\label{eq:eta_A_v_squared_diff}
\end{equation}
\textbf{Proof: }\textnormal{we can use Proposition~\ref{eq:eta_pm_product} to rewrite the difference, and the result follows.}
\begin{subequations}\nonumber
    \begin{align}
    \eta_A^2-\eta_\nu^2&=\left(\eta_+\eta_--4\nu^2\right)-\left(\eta_+\eta_--4A^2\right)=4(A^2-\nu^2)\\
    &=2(1+A^2-\nu^2)-2(1+\nu^2-A^2)=2(\eta_A-\eta_\nu)\\
    &=4(1+A^2-\nu^2)-4=4\eta_A-4\\
    &=4-(4-A^2+\nu^2)=4-4\eta_\nu
    \end{align}
\end{subequations}
\end{proposition}
\begin{proposition}
Given the products $\eta_+\eta_-$ and $\eta_A\eta_\nu$, the following identities hold:
\begin{equation}
    \eta_+\eta_-+\eta_A\eta_\nu=2(1+A^2+\nu^2).
\label{eq:eta_pm_eta_A_v}
\end{equation}
\textbf{Proof: }\textnormal{we can use Proposition~\ref{eq:eta_pm_product} and \ref{eq:eta_A_v_sum_diff} to show the result.}
\begin{subequations}\nonumber
    \begin{align}
        \eta_+\eta_-+\eta_A\eta_\nu&=\eta_A^2+4\nu^2+\eta_A\eta_\nu=\eta_A(\eta_A+\eta_\nu)+4\nu^2\\&=2\eta_A+4\nu^2=2(1+A^2+\nu^2)
    \end{align}
\end{subequations}
\end{proposition}
\begin{proposition}
The partial derivative with respect to $\nu$ of $\eta_+$, $\eta_-$, and $\eta_A$ and the product $\eta_+\eta_-$ are given by
\begin{equation}
    \frac{\partial\eta_\pm}{\partial\nu}=2\nu\pm2A,\quad\quad
    \frac{\partial\eta_A}{\partial\nu}=-2\nu,\quad\quad
    \frac{\partial\eta_\nu}{\partial\nu}=2\nu,\quad\quad
    \frac{\partial(\eta_+\eta_-)}{\partial\nu}=4\nu\eta_\nu.
\label{eq:eta_derivatives}
\end{equation}
\textbf{Proof: }\textnormal{the proof is straightforward by deriving the definitions and Proposition~\ref{eq:eta_pm_product}.}
\begin{subequations}\nonumber
\label{eq:part_derv}
    \begin{align}
        &\frac{\partial\eta_+}{\partial\nu}=\frac{\partial}{\partial\nu}\left(1+(\nu\pm A)^2\right)=2\nu\pm2A\\
        &\frac{\partial\eta_A}{\partial\nu}=\frac{\partial}{\partial\nu}\left(1+A^2-\nu^2\right)=-2\nu\\
        &\frac{\partial\eta_\nu}{\partial\nu}=\frac{\partial}{\partial\nu}\left(1+\nu^2-A^2\right)=2\nu\\
        &\frac{\partial(\eta_+\eta_-)}{\partial\nu}=\frac{\partial}{\partial\nu}\left(\eta_v^2+4A^2\right)=2\eta_\nu\frac{\partial\eta_\nu}{\partial\nu}=4\nu\eta_\nu
    \end{align}
\end{subequations}
\end{proposition}
\newrefcontext[labelprefix=S]
\printbibliography[title={Supplementary references}]
\end{refsection}

\end{document}